\documentclass[aps,preprint,nofootinbib,a4paper,11pt,longbibliography]{revtex4-1}

\usepackage{amsmath} 
\usepackage{graphicx} 
\usepackage{amsthm}
\usepackage{amssymb} 
\usepackage{yfonts}
\usepackage{hyperref}
\usepackage{footmisc}

\usepackage{MnSymbol}
\usepackage{hyperref}
\usepackage{xfrac}
\usepackage{subfig}
\usepackage{booktabs}
\hypersetup{colorlinks=true,linkcolor=blue,citecolor=red,urlcolor=blue}

\newcommand{\be}{\begin{equation}}
\newcommand{\ee}{\end{equation}}
\newcommand{\bea}{\begin{eqnarray}}
\newcommand{\eea}{\end{eqnarray}}

\newcommand{\nn}{\nonumber\\}

\def\tp{\tau_{\Pi}}

\def\CD{\mathcal{D}}

\def\CG{\mathcal{G}}

\def\CK{\mathcal{K}}
\def\CL{\mathcal{L}}

\def\CN{\mathcal{N}}

\def\CO{\mathcal{O}}

\def\CW{\mathcal{W}}

\def\lgb{\lambda_{GB}}

\begin{document}
\title{On the universal identity in second order hydrodynamics}
\author{S.~Grozdanov}
\email{grozdanov@lorentz.leidenuniv.nl}
\affiliation{Rudolf Peierls Centre for Theoretical Physics, \\University of Oxford, 1 Keble Road, \\ Oxford OX1 3NP, United Kingdom    \\and\\ Instituut-Lorentz for Theoretical Physics, \\Leiden University, Niels Bohrweg 2, \\Leiden 2333 CA, The Netherlands}
\author{A.~O.~Starinets}
\email{andrei.starinets@physics.ox.ac.uk}
\affiliation{Rudolf Peierls
Centre for Theoretical Physics,
University of Oxford, 1 Keble Road, Oxford OX1 3NP, United Kingdom\\
\vspace{1cm}
}
\preprint{OUTP-14-20P}

\begin{abstract}
We compute the 't Hooft coupling correction to the infinite coupling expression for the second order 
transport coefficient $\lambda_2$ in ${\cal N}=4$ $SU(N_c)$ supersymmetric Yang-Mills theory at finite temperature in the limit of infinite $N_c$, which originates from the $R^4$ terms in the low energy effective action of the dual type IIB string theory. Using this result, we 
show that the identity involving  the three second order transport 
coefficients, $2 \eta \tau_\Pi - 4 \lambda_1 - \lambda_2 =0$, previously shown by Haack and Yarom to
 hold universally in relativistic conformal 
field theories with string dual descriptions
to leading order in supergravity approximation, holds also at next to leading order in this theory. We also compute corrections to transport 
coefficients in a (hypothetical) strongly interacting conformal  fluid arising from the generic curvature squared terms in the 
corresponding dual gravity action (in particular, Gauss-Bonnet action), 
and show that the identity holds to linear order in the higher-derivative couplings. We discuss potential implications 
of these results for the near-equilibrium entropy production rate at strong coupling. 
\end{abstract}

\maketitle

\section{Introduction}
Fluid dynamics is currently understood as an effective theory approximating a given microscopic theory in the long-wavelength, long-time regime via a systematic derivative expansion \cite{forster}, \cite{Baier:2007ix}, \cite{Son:2007vk}, \cite{Schaefer:2014awa}. The corresponding equations of motion (Navier-Stokes equations, Burnett equations, and their generalizations) follow from combining the equations expressing conservation laws of the microscopic theory with the 
constitutive relations at a given order in the derivative expansion.\footnote{In this paper, we ignore issues related to non-analytic contributions to correlation functions at small frequency and the breakdown of the derivative expansion \cite{Kovtun:2003vj}, \cite{Kovtun:2011np}. This is justified as long as we work within classical (i.e. not quantum) gravity approximation. For  ${\cal N}=4$ $SU(N_c)$ supersymmetric Yang-Mills theory, this means staying in the limit of infinite $N_c$.}

In the simplest case of a  relativistic neutral conformal fluid in a $d$-dimensional curved spacetime, the derivative expansion of the stress-energy tensor's expectation value has the form
\begin{align}
T^{ab} = \varepsilon u^a u^b + P \Delta^{ab} + \Pi^{ab} + \cdots\,,
\end{align}  
where $u^a$ is the velocity of the fluid, $\varepsilon$ is the energy density, $P$ is the pressure fixed by the conformal invariance to obey the equation of state $\varepsilon = (d-1) P$, and the tensor $\Pi^{ab}$ involving first and second derivatives of velocity is given by 
\begin{align}
\label{disspart}
\Pi^{ab} =&~ -\eta \sigma^{ab} + \eta \tp \left[ {}^{\langle}D\sigma^{ab\rangle} + \frac{1}{d-1} \sigma^{ab} \left(\nabla\cdot u\right) \right] + \kappa \left[ R^{\langle ab \rangle} - (d-2)u_c R^{c \langle ab \rangle d} u_d  \right]  \nn
&+\lambda_1 \sigma^{\langle a}_{~~c} \sigma^{b\rangle c}  +\lambda_2 \sigma^{\langle a}_{~~c} \Omega^{b\rangle c}  +\lambda_3 \Omega^{\langle a}_{~~c} \Omega^{b\rangle c},
\end{align}
where $D \equiv u^a \nabla_a$. We use notations and sign conventions of Ref.~\cite{Son:2007vk,Baier:2007ix}, where the 
definitions of tensor structures such as vorticity $\Omega^{ab}$  appearing in Eq.~(\ref{disspart}) can also be found. 

The six transport coefficients $\eta$, $\tau_\Pi$, $\kappa$, $\lambda_1$, $\lambda_2$ and $\lambda_3$
 in Eq.~(\ref{disspart}) are determined by the underlying microscopic theory. For conformal theories at finite temperature and zero chemical potential, they scale with the appropriate power of temperature (fixed by their scaling dimension) and may depend on coupling constants and other parameters of the theory such as the rank of the gauge group. For some theories, transport coefficients have been computed in the regime of weak coupling using kinetic theory approach and in the regime of strong coupling using gauge-string duality methods  \cite{original1,original2,original3,original4}. 

Of particular interest are the properties of transport coefficients universal for all or at least some class of theories. For example, 
the dimensionless ratio of shear viscosity to entropy density exhibits such a universality: assuming validity of gauge-string duality, 
one can prove that the ratio 
is equal to $1/4\pi$ for a large class of theories in the limit described by a dual supergravity 
(usually, in the limit of infinite coupling and infinite rank of the gauge group) \cite{Kovtun:2004de}, \cite{Buchel:2004qq}, \cite{Kovtun:2003wp,Buchel:2003tz,Starinets:2008fb}. This result is very robust and holds for any quantum field theory (conformal or not) 
with a gravity dual description (see \cite{Cremonini:2011iq} for a recent summary and discussion). Coupling constant corrections to 
the viscosity-entropy ratio are not expected to be universal: in each theory, the ratio is a non-trivial function of the coupling and other parameters. In particular, in the finite-temperature ${\cal N}=4$ $SU(N_c)$ supersymmetric Yang-Mills (SYM) theory in $d=3+1$ dimensions in the limit of infinite $N_c$ and infinite 't Hooft coupling $\lambda = g^2_{YM}N_c$, the shear viscosity to entropy density ratio appears to be a monotonic function of the coupling \cite{Kovtun:2004de}, with the correction to the universal infinite coupling result being positive  \cite{Buchel:2004di,Buchel:2008sh}:
\begin{equation}
\frac{\eta}{s} = \frac{1}{4\pi} \left( 1 + 15\zeta (3) \lambda^{-3/2}+ \ldots \right)\,.
\end{equation}
Subsequent calculations revealed that in other (hypothetical) quantum field theories the corrections coming from higher derivative terms in the dual action can have either sign \cite{Kats:2007mq, Brigante:2007nu}. In particular, for a (hypothetical) field theory dual to Einstein gravity with Gauss-Bonnet higher derivative terms\footnote{As is well known \cite{Brigante:2008gz,Buchel:2009tt,Buchel:2009sk,Buchel:2010wf,Camanho:2014apa}, such a field theory would be suffering from inconsistencies such as causality violation unless restrictions are imposed on the Gauss-Bonnet coupling or other degrees of freedom are added to the Gauss-Bonnet gravity. Our working assumption is that it is possible to cure the problems in the ultraviolet without affecting the hydrodynamic (infrared) regime.}
\begin{align}\label{GBaction}
S_{GB} = \frac{1}{2\kappa_5^2} \int d^5 x \sqrt{-g} \left[ R  + \frac{12}{L^2} + \frac{\lgb}{2} L^2 \left( R^2 - 4 R_{\mu\nu} R^{\mu\nu} + R_{\mu\nu\rho\sigma} R^{\mu\nu\rho\sigma} \right) \right],
\end{align}
where $L$ is the AdS radius, one finds \cite{Brigante:2007nu}
\begin{equation}
\frac{\eta}{s} = \frac{1 - 4 \lgb}{4\pi}
\label{gbviscosity}
\end{equation}
non-perturbatively in the Gauss-Bonnet coupling $\lgb$. 

Much less is known about bulk viscosity\footnote{For conformal theories, bulk viscosity is zero.} 
\cite{Benincasa:2005iv}, \cite{Benincasa:2005qc}, \cite{Mas:2007ng}. 
A proposal for a universal inequality  involving bulk viscosity at strong coupling has been made by 
Buchel  \cite{Buchel:2007mf} but it seems there exist counterexamples to it \cite{Buchel:2011uj}. 

Universal behavior is also known to exist for second order transport coefficients. Following the observation made 
in \cite{Erdmenger:2008rm}, Haack and Yarom \cite{Haack:2008xx} showed that for relativistic conformal fluids with $U(1)$ charges
in $d>3$ space-time dimensions\footnote{In lower dimension, the coefficient $\lambda_1$ is undefined.} in the limit 
described by a dual two-derivative gravity, the following linear combination vanishes: 
\begin{equation}
 H \equiv 2 \eta \tp - 4\lambda_1 - \lambda_2  = 0\,.
 \label{universal_identity}
\end{equation}
We expect $H$ to be (generically) a non-trivial function of the coupling: perturbative analysis in QED and other theories \cite{York:2008rr} suggests that $H\neq 0$ at weak coupling. In conformal kinetic theory one finds 
$2  \eta  \tau_\Pi  +  \lambda_2 = 0$ \cite{Baier:2007ix,York:2008rr,Betz:2008me}. In the same regime, the ratio $\lambda_1/\eta \tau_\Pi$ is expected to be relatively close to 1 but the prediction $\lambda_1 = \eta \tau_\Pi$  \cite{Betz:2008me} is understood to be an artifact of a too restrictive {\it ansatz} choice for the collision integral \cite{York:2008rr}. (If the prediction were true, we would have $H=0$ in the kinetic regime.) 

For the conformal theory dual to Gauss-Bonnet gravity (\ref{GBaction}), the statement that $H$ is not identically zero 
would imply that generically one may expect $H=O\left(\lgb \right)$.

Intriguingly, Shaverin and Yarom found that $H=0$ still holds in this theory to linear order in $\lgb$ \cite{Shaverin:2012kv}. However, the universal relation does not hold to order $\lgb^2$  \cite{Grozdanov:2015asa,Grozdanov:2016fkt}\footnote{This has been independently 
found in \cite{Shaverin-thesis}, \cite{Shaverin:2015vda} via fluid-gravity methods. 
We thank E.~Shaverin and A.~Yarom for sharing these results.}. Indeed, as will be shown elsewhere \cite{Grozdanov:2015asa,Grozdanov:2016fkt}, the full non-perturbative expression for $H (\lgb)$ implies that the identity (\ref{universal_identity}) holds to linear order in $\lgb$ but is violated at $O(\lgb^2)$:
\begin{align}
\label{SecondOrderUniViolation}
H(\lgb) = - \frac{\eta}{\pi T} \frac{\left(1-\gamma_{GB}\right) \left(1 - \gamma_{GB}^2 \right) 
\left(3+2\gamma_{GB}\right)}{\gamma_{GB}^2}  = - \frac{40 \lgb^2 \eta}{\pi T}\, + \CO\left(\lgb^3\right)\,,
\end{align}
where $\gamma_{GB} = \sqrt{1-4\lgb}$. Curiously, in that theory $H(\lgb) \leq 0$ for all values of $\lgb \in (-\infty,1/4]$.

In this paper, we show that the identity $H=0$ holds in ${\cal N}=4$ SYM at next to leading order in the strong coupling expansion (in 
the limit $N_c \rightarrow \infty$). In this limit, the shear viscosity and the second order transport coefficients of  ${\cal N}=4$ SYM are given by
\begin{align}
\eta &= \frac{\pi}{8} N^2_c T^3\, \left( 1 + 135 \gamma  + \ldots \,  \right)\,, \label{tcsym1} \\
\tau_{\Pi}  &= \frac{ \left( 2 - \ln{2}\right)}{2\pi T}   + \frac{375 \gamma}{4 \pi T} \, +\ldots \,, \label{tcsym2} \\
\kappa &= \frac{N_c^2 T^2}{8}  \left( 1 - 10 \gamma  + \ldots \,  \right)\,, \label{tcsym3} \\  
\lambda_1 &=  \frac{N_c^2 T^2}{16}  \left( 1 + 350 \gamma  + \ldots \,  \right)\,,  \label{tcsym4}  \\
\lambda_2 &=- \frac{N_c^2 T^2}{16} \left( 2\ln{2}   + 5 \left(97+54 \ln{2}\right) \gamma + \ldots \,\right), \label{tcsym5} \\
\lambda_3 &=  \frac{ 25 N_c^2 T^2}{2} \, \gamma +\ldots \,,\label{tcsym6}
\end{align}
where $\gamma =  \lambda^{-3/2} \zeta (3)/8$.
%
To leading order in the strong coupling limit (i.e. at $\gamma\rightarrow 0$), the results (\ref{tcsym1}) - (\ref{tcsym6}) were obtained in \cite{Policastro:2001yc,Baier:2007ix,Bhattacharyya:2008jc} using gauge-gravity and fluid-gravity dualities. Coupling constant corrections to all the coefficients except $\lambda_2$ were previously computed from the higher-derivative terms in the low-energy effective action of type IIB string theory \cite{Buchel:2004di, Benincasa:2005qc, Buchel:2008sh, Buchel:2008ac,Buchel:2008bz,Buchel:2008kd,Saremi:2011nh}. The $O\left(\lambda^{-3/2}\right)$ correction in the expression for $\lambda_2$ is the new result obtained in Section \ref{sec:tHooftN4} of the present paper.

The corrections in formulae (\ref{tcsym1}) - (\ref{tcsym6}) can be trusted so long as they remain (infinitesimally) small relative to the leading order ($\lambda\rightarrow \infty$) result, as they are obtained by treating the higher-derivative terms in the equations of motion perturbatively. To leading order in the strong coupling limit, the coefficients  (\ref{tcsym1}) - (\ref{tcsym6})  are independent of the coupling, in sharp contrast with their weak coupling behavior \cite{Huot:2006ys}. The coefficient $\lambda_3$ vanishes at $\lambda\rightarrow \infty$, and was argued to vanish also at $\lambda \rightarrow 0$ (this appears to be a generic property of weakly coupled theories). The full coupling constant dependence of transport coefficients (even at infinite $N_c$) appears to be beyond reach. 

The results (\ref{tcsym1}), (\ref{tcsym2}), (\ref{tcsym4}) and  (\ref{tcsym5})  imply that the identity (\ref{universal_identity}) holds in  ${\cal N}=4$ SYM at the order $O\left(\lambda^{-3/2}\right)$ in the strong coupling expansion.

\section{Coupling constant correction to the second order transport coefficient $\lambda_2$ in $\CN=4$ SYM theory}
\label{sec:tHooftN4}

Coupling constant corrections to transport coefficients in $\CN=4$ SYM can be computed using the 
following dual five-dimensional gravitational action with the $R^4$ higher derivative term\footnote{As argued in \cite{Buchel:2008ae}, 
to compute physical quantities in the hydrodynamic regime of field theories dual to ten-dimensional type IIB supergravity with five compact dimensions, it is sufficient to consider only the reduced five-dimensional action.} 
\begin{align}
S = \frac{1}{2\kappa_5^2} \int d^5 x \sqrt{-g} \left(R  + \frac{12}{L^2} + \gamma \CW \right),
\label{hd-action}
\end{align}
where $\gamma = \alpha'^3 \zeta(3) / 8$ which is related to the value of the 't Hooft coupling $\lambda$ in $\CN=4$ SYM 
via $\alpha' / L^2 = \lambda^{-1/2}$. We set the AdS radius $L=1$ in the following. The effective five-dimensional gravitational 
constant is connected to the rank of the gauge group by $\kappa_5 = 2\pi /N_c$. The term $\CW$ is given in terms of the Weyl tensor $C_{\mu\nu\rho\sigma}$ by
\begin{align}
\CW = C^{\alpha\beta\gamma\delta}C_{\mu\beta\gamma\nu} C_{\alpha}^{~\rho\sigma\mu} C^{\nu}_{~\rho\sigma\delta} + \frac{1}{2} C^{\alpha\delta\beta\gamma} C_{\mu\nu\beta\gamma} C_{\alpha}^{~\rho\sigma\mu} C^\nu_{~\rho\sigma\delta}.
\end{align}
The $\alpha'$-corrected black brane solution corresponding to the action (\ref{hd-action}) was found in \cite{Gubser:1998nz}:
\begin{align}
ds^2 = \frac{r_0^2}{u} \left( - f(u) Z_t dt^2 + dx^2 +dy^2 +dz^2 \right) + Z_u \frac{du^2}{4u^2 f}\,,
\label{corrected_metric}
\end{align}
where $f(u) = 1 - u^2$, $r_0$ is the parameter of non-extremality of the black brane geometry,
and the functions $Z_t$ and $Z_u$ are given by
\begin{align}
Z_t = 1 - 15\gamma\left(5u^2+5u^4-3 u^6 \right) , && Z_u = 1 + 15\gamma \left(5u^2 + 5 u^4 - 19 u^6 \right) . 
\end{align}
The Hawking temperature corresponding to the metric (\ref{corrected_metric}) is $T = r_0 (1+15\gamma)/\pi$. The standard black three-brane solution is recovered in the limit $\gamma\rightarrow 0$.

To compute the 't Hooft coupling correction to the second order transport coefficient $\lambda_2$, we use the method 
of three-point functions\footnote{Explicit holographic calculations of the equilibrium real-time three-point and four-point functions in strongly 
coupled ${\cal N}=4$ SYM at finite temperature have been pioneered in \cite{Barnes:2010jp}, \cite{Barnes:2010ev}, \cite{Arnold:2011hp}.} 
and the associated Kubo formulae developed by Moore, Sohrabi and Saremi  \cite{Moore:2010bu,Saremi:2011nh} and by Arnold, Vaman, Wu and Xiao
\cite{Arnold:2011ja}. The relevant retarded three-point functions of the stress-energy tensor are defined in the Schwinger-Keldysh closed time path formalism \cite{Schwinger:1960qe,Keldysh:1964ud}. We now review the key elements of the method. 

Consider a theory described by a microscopic Lagrangian $\CL \left[\phi, h\right]$, where $\phi$ collectively denotes matter fields and $h$ corresponds to a metric perturbation of a fixed background $g$ (all tensor indices are suppressed). The degrees of freedom of the theory are then doubled, $\phi \to \phi^\pm$, $g \to g^\pm$, $h \to h^\pm$, where we use the index $\pm$ to denote whether the fields are defined on a ``$+$"-time contour running from $t_0$ towards the final time $t_f > t_0$, or the ``$-$"-contour with time running from the future $t_f$ backwards to $t_0$. For the field theory considered at a finite temperature $T = 1/\beta$, the two separated real time contours can be joined together by a third, imaginary time part of the contour running between $ t_f$ and $t_f - i \beta$. We use $\varphi$ to denote fields defined in the Euclidean theory on the imaginary time contour. The generating functional of the stress-energy tensor correlation functions can then be written as
\begin{align}
W\left[h^+, h^-\right] =& \ln \int \CD\phi^+\CD\phi^-\CD\varphi \exp \left\{i\int d^4x^+ \sqrt{-g^+} \CL\left[\phi^+(x^+), h^+\right] \right.  \nn
& \left.  - \int_0^\beta d^4 y \CL_E \left[\varphi(y)\right] - i \int d^4 x^- \sqrt{-g^-} \CL\left[\phi^-(x^-), h^-\right] \right\}.
\end{align}
It is convenient to introduce the Keldysh basis $\phi_R = \frac{1}{2} \left( \phi^+ + \phi^- \right)$ and $\phi_A = \phi^+ - \phi^-$, and similarly for the metric perturbation and the stress-energy tensor. After variation, the classical expectation values always obey $\phi^+ = \phi^-$, hence all fields with an index $A$ will vanish and one can define $T^{ab} \equiv T^{ab}_R$. Explicitly, 
\begin{align}
\left\langle T^{ab}_R (x) \right\rangle = -\frac{2i}{\sqrt{-g}} \frac{\delta W }{ \delta h_{A\;ab}(x) } \biggr|_{h = 0}.
\end{align}
The expectation value of $T_R$ at $x=0$ can then be expanded as 
\begin{align}
\left\langle T^{ab}_R (0) \right\rangle =& G^{ab}_R (0) - \frac{1}{2} \int d^4 x G^{ab,cd}_{RA} (0,x) h_{cd}(x)   \nn
& + \frac{1}{8} \int d^4xd^4y G^{ab,cd,ef}_{RAA}(0,x,y) h_{cd}(x)h_{ef}(y) + \ldots,
\end{align}
where $G_{RAA...}$ denote the {\it fully retarded} Green's functions \cite{Wang:1998wg} obtained by taking the appropriate number of derivatives with respect to   $h_A$ and $h_R$ \cite{Moore:2010bu}
\begin{align}
G^{ab,cd,\ldots}_{RA\ldots} (0, x, \ldots) = \frac{(-i)^{n-1} (-2 i)^n \delta^n W   }{\delta h_{A\;ab} (0) \delta h_{R\;cd} (x) \ldots } \biggr|_{h=0}  = (-i)^{n-1} \left\langle T_R^{ab} (0) T_A^{cd}(x) \ldots  \right\rangle\,.
\end{align}
Denoting the space-time coordinates of the four-dimensional field theory by $t,x,y,z$ and choosing the momentum along the $z$ axis, one can write the following Kubo formula for the coefficient $\lambda_2$ \cite{Moore:2010bu,Saremi:2011nh}
\begin{align}\label{L2KuboThreePt}
\lambda_2 = 2\eta\tp - 4 \lim_{p,q\to 0} \frac{\partial^2}{\partial p^0 \partial q^z} G^{xy,ty,xz}_{RAA} (p,q)\,.
\end{align}

The three-point functions are calculated by solving the bulk equations of motion  to second order in metric perturbations of the background $g_{\mu\nu}^{(0)}$ (\ref{corrected_metric}),
\begin{align}
g_{\mu\nu}  = g_{\mu\nu}^{(0)} + \epsilon \frac{r_0^2}{u}  h^{(1)}_{\mu\nu} + \epsilon^2  \frac{r_0^2}{u} h^{(2)}_{\mu\nu},
\end{align}
where $\epsilon$ serves as a book-keeping parameter indicating the order of the perturbation. 
We impose the Dirichlet condition 
$h^{(2)}_{\mu\nu} = 0$ at the boundary \cite{Saremi:2011nh}. The three-point functions are found by taking functional derivatives 
of the on-shell action with respect to the boundary value  $h^{(b)}_{\mu\nu} = h^{(1)}_{\mu\nu} \left(u\to 0\right)$. 
A simplifying feature of this procedure is that since equations of motion are solved to order $\epsilon^2$, only the boundary term contributes to the three-point function and hence no bulk-to-bulk propagators appear in the calculation. To compute $G^{xy,ty,xz}_{RAA} (p,q)$, we turn on the following set of metric perturbations
\begin{align}
h_{xy} = h_{xy} (r) e^{- i p^0 t + i q^z z }, & & h_{xz} = h_{xz} (r) e^{-i p^0  t }, & & h_{ty} = h_{ty} (r) e^{i q^z  z }. \label{Pol3}
\end{align}
In the following, we use the notations $\omega \equiv p^0$, $q \equiv q^z$ and $T_0 \equiv r_0/\pi$. 

At first order in $\epsilon$, the metric perturbations can be written as expansions in $\gamma$ 
\begin{align}\label{N4FirstOrderFluct}
&h^{(1)}_{xy} = h_{xy}^{(b)} e^{-i \omega t + i q z} \left( Z_{xy} + \gamma Z^{(\gamma)}_{xy} \right) , \\
&h^{(1)}_{xz} = h_{xz}^{(b)}  e^{-i \omega t} \left( Z_{xz} + \gamma Z^{(\gamma)}_{xz} \right), \\ 
&h^{(1)}_{ty} = h_{ty}^{(b)}  e^{i q z} \left( Z_{ty} + \gamma Z^{(\gamma)}_{ty} \right)\,,
\end{align} 
where $Z^{(\gamma)}_{xy}$, $Z^{(\gamma)}_{xz}$ and $Z^{(\gamma)}_{ty}$ are treated as perturbations (in $\gamma$) of the main solutions $Z_{xy}$, $Z_{xz}$ and $Z_{ty}$ .
%
%
The equations of motion for the metric fluctuations follow from the action  (\ref{hd-action}), where the $\gamma$-dependent part is treated as a perturbation. The differential equation for $Z_{xy}$  is 
\begin{align}\label{ZxyEq}
\partial_u^2 Z_{xy} - \frac{1+u^2}{u(1-u^2)} \partial_u Z_{xy}+ \frac{\omega^2 - q^2(1-u^2)}{4\pi^2 T_0^2 u (1 - u^2)^2} Z_{xy} = 0.
\end{align}
The functions  $Z_{xz}$ and $Z_{ty}$ obey the same differential equation \eqref{ZxyEq} with $q$ and $\omega$, respectively, 
set to zero. Note that we cannot impose the incoming wave boundary condition at the horizon on $Z_{ty}$, 
as it has no time dependence. Instead, the Dirichlet condition $Z_{ty} = 0$ is used \cite{Saremi:2011nh}.  

The solutions to quadratic order in $\omega$ and $q$ are given by
\begin{align}
Z_{xy} =\,\,&  \left(1-u^2\right)^{-\frac{i \omega }{4 \pi  T_0} (1-15 \gamma )}\,  \biggr[1 \, + \nn
&    +   \frac{6 \ln  (u+1) \left[ \omega^2 \ln \left(\frac{u+1}{4}\right)+4 \omega^2-4  q^2\right]  +   \omega^2 \left[  \pi ^2-6 \ln ^2(2)      -12 \omega^2 \, \text{Li}_2\left(\frac{1-u}{2}\right) \right]}{96 \pi^2  T_0^2}     \biggr],  \\
Z_{xz} =\,\,& \left(1-u^2\right)^{-\frac{i \omega }{4 \pi  T_0} (1-15 \gamma )}\,  \biggr[1 \, + \nn
&    +   \frac{6 \omega^2 \ln  (u+1) \left[  \ln \left(\frac{u+1}{4}\right)+4 \right]  +   \omega^2 \left[  \pi ^2-6 \ln ^2(2)      -12 \omega^2 \, \text{Li}_2\left(\frac{1-u}{2}\right) \right]}{96 \pi^2
  T_0^2}     \biggr] , \\
Z_{ty} = \,\,&   1 - u^2 - \frac{q^2 u ( 1- u)}{4 \pi ^2 T_0^2} .
\end{align}
We now use these solutions in the full equations of motion to find the differential equations obeyed by $Z^{(\gamma)}_{xy}$, $Z^{(\gamma)}_{xz}$ and $Z^{(\gamma)}_{ty}$. All three equations have the form (indices suppressed)
\begin{align}
\partial^2_u Z^{(\gamma)} - \frac{1+u^2}{u(1-u^2)} \partial_u Z^{(\gamma)}+ \frac{\omega^2 - q^2(1-u^2)}{4\pi^2 T_0^2 u (1 - u^2)^2} Z^{(\gamma)} = \CG(u),
\end{align}
where the functions $\CG(u)$ on the right hand side are, respectively,
\begin{align}
\CG_{xy} =\,\,& h_{xy}^{(b)} \left(1-u^2\right)^{-\frac{i \omega}{4 \pi  T_0}}  \Biggr[  \frac{3 i \omega u^2 \left(129 u^4+94 u^2-25\right)}{\pi  T_0 \left(u^2-1\right)}  \nn
&+  \frac{1}{4 \pi ^2 T_0^2 (u-1) (u+1)^2}  \biggr[   -6 \omega^2 u^2 (u+1) \left(129 u^4+94 u^2-25\right) \ln (u+1) \nn
&+6 \omega^2 u^2
   \left(-129 u^5+35 u^3-89 u^2+(u+1) \left(129 u^4+94 u^2-25\right) \ln (2)+30
   u+30\right)  \nn
&+q^2 u (u+1) \left(774 u^5-1625 u^4+564 u^3+225 u^2-150 u+75\right) \biggr] \Biggr], 
\end{align}
\begin{align}
\CG_{xz} =\,\,& h_{xz}^{(b)} \left(1-u^2\right)^{-\frac{i \omega}{4 \pi  T_0}}  \Biggr[  \frac{3 i \omega u^2 \left(129 u^4+94 u^2-25\right)}{\pi  T_0 \left(u^2-1\right)}  \nn
&+  \frac{1}{4 \pi ^2 T_0^2 (u-1) (u+1)^2}  \biggr[   -6 \omega^2 u^2 (u+1) \left(129 u^4+94 u^2-25\right) \ln (u+1) \nn
&+6 \omega^2 u^2
   \left(-129 u^5+35 u^3-89 u^2+(u+1) \left(129 u^4+94 u^2-25\right) \ln (2)+30
   u+30\right)  \biggr] \Biggr],
\end{align}
\begin{align}
\CG_{ty} =\,\,& h_{ty}^{(b)} \Biggr[ 720 u^4 \left(4-3 u^2\right)+ \frac{5 q^2 u \left(432 u^5+551 u^4-576 u^3+15 u^2+15\right)}{4 \pi ^2
   T_0^2} \Biggr].
\end{align}
The solutions 
(to quadratic order in $\omega$ and $q$) are given by 
\begin{align}
Z^{(\gamma)}_{xy} =\,\,& \left(1-u^2\right)^{-\frac{i \omega}{4 \pi  T_0}}   
\Biggr\{ \frac{i \omega u^2 \left(43 u^4+135 u^2+195\right)}{4 \pi  T_0}  +\frac{1}{48 \pi ^2 T_0^2} 
    \,  \biggr[180
   \omega^2 \text{Li}_2\left(\frac{1-u}{2}\right)  \nn
& -258 \omega^2 u^6+258
   \omega^2 u^5-810 \omega^2 u^4 -160 \omega^2 u^3 -1170 \omega^2
   u^2       \nn
&   +6 \omega^2 \left(43 u^4+135 u^2+195\right) u^2 \ln
   \left(\frac{2}{u+1}\right)   \nn
&+3630 \omega^2 \ln (u+1)+30 \omega^2 \ln (64) \ln (u+1) -15 \pi ^2 \omega^2+90 \omega^2 \ln ^2(2) \nn
&+258 q^2 u^6-780 q^2 u^5+810 q^2 u^4  \nn
&-1000 q^2 u^3+1170 q^2 u^2-2100 q^2 u+2100 q^2 \ln
   (u+1)      \biggr]\Biggr\}  , \\
   %
Z^{(\gamma)}_{xz} = \,\,& \left(1-u^2\right)^{-\frac{i \omega}{4 \pi  T_0}}   \Biggr\{ \frac{i \omega u^2 \left(43 u^4+135 u^2+195\right)}{4 \pi  T_0}  +\frac{1}{48 \pi ^2 T_0^2}     \,  \biggr[180
   \omega^2 \text{Li}_2\left(\frac{1-u}{2}\right)  \nn
& -258 \omega^2 u^6+258
   \omega^2 u^5-810 \omega^2 u^4 -160 \omega^2 u^3 -1170 \omega^2
   u^2       \nn
&   +6 \omega^2 \left(43 u^4+135 u^2+195\right) u^2 \ln
   \left(\frac{2}{u+1}\right)   \nn
&+3630 \omega^2 \ln (u+1)+30 \omega^2 \ln (64) \ln (u+1) -15 \pi ^2 \omega^2+90 \omega^2 \ln ^2(2)  \biggr]\Biggr\} \,, \\
%
Z^{(\gamma)}_{tz} = \,\,&-15 \left(5 u^2-8 u^6 + 3 u^8\right) - \frac{5 q^2 u^2 \left(1+24 u^4 -16 u^5 - 9 u^6\right)}{4 \pi ^2 T_0^2}.
\end{align}

The next step is to use the above solutions to find the second-order perturbation $h^{(2)}_{xy}$ of the metric to linear order 
in $\gamma$ and to quadratic order in $\omega$ and $q$. We begin by computing the action (\ref{hd-action}) with
\begin{align}
&g_{xy} = g_{xy}^{(0)} + \frac{(\pi T_0)^2}{u} \epsilon h^{(1)}_{xy} + \frac{(\pi T_0)^2}{u} \epsilon^2 h^{(2)}_{xy},  \\
&g_{xz} = g_{xz}^{(0)} + \frac{(\pi T_0)^2}{u} \epsilon h^{(1)}_{xz}, \\
&g_{ty} = g_{ty}^{(0)} + \frac{(\pi T_0)^2}{u} \epsilon h^{(1)}_{ty}
\end{align}
 to order $\epsilon^4$. This gives us the effective action and the equation of motion for the fluctuation $h^{(2)}_{xy}$  that can be solved perturbatively to linear order in $\gamma$ and quadratic order in $\omega$ and $q$. We can look for a  solution in the form 
\begin{align}
h^{(2)}_{xy} = h^{(b)}_{xz} h^{(b)}_{ty} e^{-i\omega t + i q z} \left( Y_{xy} + \gamma Y^{(\gamma)}_{xy} \right).
\end{align}
At $\gamma=0$, the full fluctuation equation is
\begin{align}
 \partial^2_u Y_{xy}    -   \frac{1+u^2}{u \left(1-u^2\right)}  \partial_u Y_{xy} +\frac{\omega^2 - q^2(1-u^2)}{4\pi^2 T_0^2 u (1 - u^2)^2} Y_{xy}   -\frac{\omega \, q \, h^{(b)}_{xz} h^{(b)}_{ty}}{4 \pi ^2 T_0^2 u  \left(1-u^2\right)^2}   Z_{xz} Z_{ty} = 0,
\end{align}
where only the $\omega$ and $q$-independent  parts of $Z_{xz}$ and $Z_{ty}$ are relevant for our purposes.

By further writing $Y_{xy}^{(\gamma)} = \left(1-u^2\right)^{-\frac{i \omega}{4 \pi  T_0}} y(u)$, the differential equation for $y(u)$
at quadratic order in $\omega$ and $q$ is simply 
\begin{align}
\partial^2_u y - \frac{1+u^2}{u \left(1-u^2\right)} \partial_u y +\frac{\omega q u \left(774 u^5+175 u^4+564 u^3+225 u^2-150
   u+75\right)}{4 \pi ^2 T_0^2 \left(1-u^2\right)}  = 0.
\end{align}
The full solution of the two differential equations is given by 
\begin{align}
&h^{(2)}_{xy} = h_{xz}^{(b)} h_{ty}^{(b)}  e^{-i \omega t + i q z} \, \frac{\omega q}{4\pi^2 T_0^2} \,\, \biggr[  - \left(1-u^2\right)^{-\frac{i \omega }{4 \pi  T_0} (1-15 \gamma )} \ln (u+1)   \nn
& + \gamma \left(1-u^2\right)^{-\frac{i \omega }{4 \pi  T_0} } \left[ \frac{1}{6} u \left(129 u^5+42 u^4+405 u^3+220 u^2+585 u+1110\right)-185 \ln
   (u+1) \right] \biggr].
\end{align}

We now compute the holographic stress-energy tensor for the induced metric $\gamma_{\mu\nu}$, 
\begin{align}\label{N4T}
T^{\mu\nu} = - \sqrt{-\gamma} \frac{N_c^2}{4\pi^2}  \frac{(\pi T_0 )^2}{u} \left[ K^{\mu\nu} - K \gamma^{\mu\nu} + 3 \left(\gamma^{\mu\nu} - \frac{1}{6} G^{\mu\nu}_{(\gamma)} \right) \right],
\end{align} 
which has the same tensorial form as in the Einstein-Hilbert gravity, with no higher-derivative terms contributing \cite{Saremi:2011nh}. Taking the derivatives of $T^{xy}$ with respect to the boundary values of $h^{(1)}_{xz}$ and $h^{(1)}_{ty}$, we find the three-point function,
\begin{align}
G^{xy,ty,xz}_{RAA} (p, q) = \frac{N_c^2}{16} \,p^0 q^z  T_0^2 \left(1 + 380 \gamma \right)\,,
\end{align}
and, using the Kubo formula (\ref{L2KuboThreePt}), the coefficient $\lambda_2$ in Eq.~(\ref{tcsym5}).

\section{Curvature squared corrections to second order transport coefficients}

In this Section we determine corrections to the second order transport coefficients in a (hypothetical) four-dimensional CFT dual to a bulk gravity 
with generic curvature squared terms. The five-dimensional bulk action is
\begin{align}\label{R2Th}
S_{R^2} = \frac{1}{2 \kappa_5^2 } \int d^5 x \sqrt{-g} \left[ R - 2 \Lambda + L^2 \left( \alpha_1 R^2 + \alpha_2 R_{\mu\nu} R^{\mu\nu} +
\alpha_3 R_{\mu\nu\rho\sigma} R^{\mu\nu\rho\sigma}  \right) \right],
\end{align}
where the cosmological constant $\Lambda = - 6 / L^2$ (we set $L=1$ in the rest of this Section). For generic values of  the coefficients $\alpha_1$,  
$\alpha_2$, $\alpha_3$, the curvature squared terms are treated perturbatively. The Gauss-Bonnet action is obtained from the action 
(\ref{R2Th}) by  
setting $\alpha_1 = \alpha_3 = \lambda_{GB}/2$ and $\alpha_2 = - 2 \lambda_{GB}$.

To compute curvature squared corrections to second order transport coefficients in a dual four-dimensional quantum field theory to linear order in $\alpha_i$, 
one can use the field redefinition and the already known results for  ${\cal N}=4$ SYM and Gauss-Bonnet gravity.

First, we set $\alpha_3 = 0$ and use the field redefinition discussed in \cite{Brigante:2007nu}, \cite{Kats:2007mq}\,,
\begin{align}\label{FieldRedef}
g_{\mu\nu} = \bar g_{\mu\nu} + \alpha_2 \bar R_{\mu\nu} - \frac{1}{3} \left(\alpha_2 + 2 \alpha_1 \right) \bar g_{\mu\nu} \bar R,
\end{align}
to rewrite the action 
(\ref{R2Th}) in the two-derivative form (to linear order in $\alpha_1$ and $\alpha_2$):
\begin{align}\label{SR2FD}
\tilde S \equiv S_{R^2} \left[\alpha_3 = 0 \right] =  \frac{1 + \CK}{2 \kappa_5^2 } \int d^5 x \sqrt{- \bar g} \left[ \bar R - 2 \bar \Lambda \right] + \CO (\alpha_i^2).
\end{align}
Here, in the notations of \cite{Brigante:2007nu}, $\CK = \frac{2 \Lambda}{3} \left(5\alpha_1 + \alpha_2 \right)$ and $\bar \Lambda = \frac{\Lambda}{1+\CK}$.
The field redefinition \eqref{FieldRedef} implies that the metric satisfies 
\begin{align}\label{MetricRescaling}
g_{\mu\nu} = A^2 \bar g_{\mu\nu} + \CO (\alpha_i^2),
\end{align}
where
\begin{align}\label{Adef}
A=1 - \frac{\CK}{3} + \CO (\alpha_i^2) .
\end{align}
We can further transform the metric $\bar g_{\mu\nu}$ to bring the action $\tilde S$ into the standard Einstein-Hilbert form with the cosmological constant 
$\Lambda$ dual to $\CN=4$ supersymmetric $SU(N_c)$ Yang-Mills theory in the regime of infinite 't Hooft coupling and infinite $N_c$. Indeed, 
consider a new metric $\tilde g_{\mu\nu}$ defined by $\bar g_{\mu\nu} = B^2 \tilde g_{\mu\nu}$ (the metric determinant and the 
Ricci scalar transform, correspondingly, as $\sqrt{-\bar g} = B^5 \sqrt{-\tilde g}$ and $\bar R = B^{-2} \tilde R$).
With $B$ given by
\begin{align}\label{Bdef}
B = 1+\frac{\CK}{2} + \CO (\alpha_i^2) ,
\end{align}
we can now  write the action \eqref{SR2FD} in the standard Einstein-Hilbert form,
\begin{align}\label{StildeEH}
\tilde S =  \frac{1}{2 \tilde\kappa_5^2} \int d^5 x \sqrt{-\tilde g} \left[ \tilde R - 2 \Lambda \right] +  \CO (\alpha_i^2),
\end{align}
where the redefined Newton's constant is
\begin{align}\label{KappaReDef}
\tilde\kappa_5^2 = B^{-5} \kappa_5^2.
\end{align}
The original metric $g_{\mu\nu}$, which is related to $\tilde g_{\mu\nu}$ by $g_{\mu\nu} = A^2 B^2 \tilde g_{\mu\nu} +  \CO (\alpha_i^2)$, can be written to linear order in $\alpha_1$ and $\alpha_2$ as
\begin{align}\label{ggtRel}
g_{\mu\nu} = e^{-2\omega} \tilde g_{\mu\nu} +  \CO (\alpha_i^2)\,,
\end{align}
where $e^{-2\omega} = 1 / A$.
%
%
%
From Eqs.~\eqref{StildeEH}, \eqref{ggtRel}, it is clear that the stress-energy tensor $\tilde T^{\mu\nu}_{\CN=4}$ of $\CN = 4$ SYM 
computed from \eqref{StildeEH}, with $\tilde\kappa_5^2$, is related to the stress-energy tensor of a theory dual to $S_{R^2} [\alpha_3 = 0]$ by a global Weyl transformation
\begin{align}
T^{\mu\nu}_{R^2} \left[\alpha_3 = 0\right] = e^{6\omega} \, \tilde T^{\mu\nu}_{\CN=4},
\end{align}
and the redefinition of $\kappa_5^2$  given by Eq.~\eqref{KappaReDef}. 
Then the scaling arguments\footnote{See e.g. \cite{Baier:2007ix} for a discussion of Weyl transformations in hydrodynamics.} imply
\begin{align}
\eta = e^{3\omega} \tilde \eta, && \eta\tp = e^{2\omega} \tilde\eta\tilde\tp, && \lambda_{1,2,3} = e^{2\omega} \tilde\lambda_{1,2,3},&& \kappa = e^{2\omega} \tilde\kappa,
\end{align}
where all the transport coefficients with the overhead tildes depend on $\tilde\kappa_5^2$. The $\CN=4$ SYM theory coefficients are
\begin{align}
&\tilde\eta = \frac{\pi^3 \tilde T^3}{2\tilde\kappa_5^2}, && \tilde\tp = \frac{2-\ln 2}{2\pi\tilde T},&&\tilde\kappa =  \frac{\tilde\eta}{\pi \tilde T}, \\
& \tilde\lambda_1 = \frac{\tilde\eta}{2\pi\tilde T}, &&\tilde\lambda_2 = - \frac{\tilde\eta \ln 2}{\pi\tilde T}, &&\tilde\lambda_3 = 0 .
\end{align}
The temperature of the $\CN = 4$ SYM theory is given by $\tilde T = r_+ / \pi$, where $r_+$ is the radial position of the black brane horizon. 
Using Eqs.~\eqref{Adef} and \eqref{Bdef}, we find that the shear viscosity in a (hypothetical) field theory dual to the gravitational background described by the action 
 $S_{R^2} [\alpha_3 = 0]$ is given by
\begin{align}\label{etaKP}
\eta = A^{3/2} B^5 \frac{r_+^3}{2 \kappa^2_5} = \frac{r_+^3}{2 \kappa^2_5} \Biggl( 1 - 8 \left(5 \alpha_1 + \alpha_2 \right) \Biggr) + \CO (\alpha_1^2, \alpha_2^2)\,,
\end{align}
which agrees\footnote{When comparing with  \cite{Kats:2007mq}, one should note that $z_0$ in \cite{Kats:2007mq} denotes 
the location of the horizon in the solution unaffected by curvature squared terms, whereas our $r_+$ is the horizon of the corrected solution. 
The relation between these parameters brings in the $\alpha_3$ dependence, which may appear to be missing from the full expression for the shear viscosity (\ref{etaKP2}). \label{ftn9}} with the results obtained in \cite{Kats:2007mq,Banerjee:2010zd}. Here $r_+$ is the location of the event horizon of a black brane solution to the equations of motion following from the action $S_{R^2} [\alpha_3 = 0]$.

The full result (for arbitrary $\alpha_3$ to linear order) is thus
\begin{align}\label{etaKPfull}
\eta  = \frac{r_+^3}{2 \kappa^2_5} \Biggl( 1 - 8 \left(5 \alpha_1 + \alpha_2 \right) \Biggr) + C_\eta \alpha_3 + \CO (\alpha_i^2)\,,
\end{align}
where the coefficient $C_\eta$ remains undetermined. Similarly, the second-order transport coefficients to linear order in $\alpha_1$ and $\alpha_2$ 
are given by the corresponding $\CN=4$ SYM results multiplied by $A B^5$:
\begin{align}
\label{zero_alpha}
&\eta \tp = \frac{r_+^2 \left( 2 - \ln 2\right) }{4 \kappa_5^2} \left( 1 - \frac{26}{3} \left(5 \alpha_1 + \alpha_2 \right) \right) + C_{\tp} \alpha_3 + \CO(\alpha_i^2), \\
&\kappa = \frac{r_+^2}{2 \kappa_5^2} \left( 1 - \frac{26}{3} \left(5 \alpha_1 + \alpha_2 \right) \right) + C_{\kappa} \alpha_3 + \CO(\alpha_i^2), \\
&\lambda_1 = \frac{r_+^2}{4 \kappa_5^2} \left( 1 - \frac{26}{3} \left(5 \alpha_1 + \alpha_2 \right) \right) + C_{\lambda_1} \alpha_3 + \CO(\alpha_i^2) , \\
&\lambda_2 = - \frac{r_+^2 \ln 2 }{2 \kappa_5^2} \left( 1 - \frac{26}{3} \left(5 \alpha_1 + \alpha_2 \right) \right) + C_{\lambda_2} \alpha_3 + \CO(\alpha_i^2) , \\
&\lambda_3 =  C_{\lambda_3} \alpha_3 + \CO(\alpha_i^2) .
\end{align}
Here we added the undetermined terms linear in $\alpha_3$. To restore the dependence on $\alpha_3$ (to linear order), recall that the 
Gauss-Bonnet expressions for transport coefficients would be restored (to linear order in $\lambda_{GB}$) by substituting 
$\alpha_1= \lambda_{GB}/2$, $\alpha_2 = - 2 \lambda_{GB}$ 
and $\alpha_3 = \lambda_{GB}/2$. For example, according to (\ref{zero_alpha}), 
the coefficient $\kappa$ in the holographic Gauss-Bonnet liquid to linear order in  $\lambda_{GB}$ should be equal to
\begin{align}
\label{kappa_comp}
\kappa  = \frac{r_+^2}{2 \kappa_5^2} \left( 1 - \frac{13}{3} \lambda_{GB} \right) + C_{\kappa} \frac{\lambda_{GB}}{2} + \CO(\alpha_i^2)\,.
\end{align}
On the other hand, all transport coefficients of the holographic Gauss-Bonnet liquid are known explicitly (non-perturbatively \cite{Grozdanov:2015asa,Grozdanov:2016fkt} and to linear order \cite{Shaverin:2012kv}):
\begin{align}
&\eta \tp = \frac{r_+^2 \left( 2 - \ln 2\right) }{4 \kappa_5^2} - \frac{r_+^2 \left(25 - 7 \ln 2 \right)}{8 \kappa_5^2} \lgb  + \CO(\lgb^2), \\
&\kappa = \frac{r_+^2}{2 \kappa_5^2} - \frac{17 r_+^2}{4 \kappa_5^2 } \lgb + \CO(\lgb^2), \label{compar_kappa} \\
&\lambda_1 = \frac{r_+^2}{4 \kappa_5^2} - \frac{9 r_+^2}{8\kappa_5^2} \lgb + \CO(\lgb^2), \\
&\lambda_2 = - \frac{r_+^2 \ln 2 }{2 \kappa_5^2} - \frac{7 r_+^2 \left(1 - \ln 2\right) }{4 \kappa_5^2} \lgb+ \CO(\lgb^2), \\
&\lambda_3 = - \frac{14 r_+^2}{\kappa_5^2} \lgb + \CO(\lgb^2) .
\end{align}
Comparing Eqs.~(\ref{kappa_comp}) and (\ref{compar_kappa}), and taking into account  $\alpha_3 = \lambda_{GB}/2$, we read off the coefficient 
$C_\kappa = - 25 r_+^2 /6\kappa_5^2$. All other coefficients are determined in the same way, and we find the transport coefficients of a 
(hypothetical) holographic liquid described by the dual gravitational action \eqref{R2Th} to linear order\footnote{See footnote \ref{ftn9}.} in $\alpha_i$:
\begin{align}
&\eta = \frac{r_+^3}{2 \kappa^2_5} \left( 1 - 8 \left(5 \alpha_1 + \alpha_2 \right) \right) + \CO (\alpha_i^2), \label{etaKP2} \\
&\eta \tp = \frac{r_+^2 \left( 2 - \ln 2\right) }{4 \kappa_5^2} \left( 1 - \frac{26}{3} \left(5 \alpha_1 + \alpha_2 \right) \right) - \frac{r_+^2 \left(23 + 5 \ln 2\right)}{ 12 \kappa_5^2} \alpha_3+ \CO(\alpha_i^2), \label{etatpKP} \\
&\kappa = \frac{r_+^2}{2 \kappa_5^2} \left( 1 - \frac{26}{3} \left(5 \alpha_1 + \alpha_2 \right) \right) - \frac{25 r_+^2}{6 \kappa_5^2} \alpha_3 + \CO(\alpha_i^2),\label{kappaKP} \\
&\lambda_1 = \frac{r_+^2}{4 \kappa_5^2} \left( 1 - \frac{26}{3} \left(5 \alpha_1 + \alpha_2 \right) \right) - \frac{r_+^2}{12 \kappa_5^2} \alpha_3 + \CO(\alpha_i^2) ,\label{l1KP} \\
&\lambda_2 = - \frac{r_+^2 \ln 2 }{2 \kappa_5^2} \left( 1 - \frac{26}{3} \left(5 \alpha_1 + \alpha_2 \right) \right) - \frac{r_+^2 \left(21 + 5 \ln 2\right)}{6\kappa_5^2} \alpha_3 + \CO(\alpha_i^2) ,\label{l2KP} \\
&\lambda_3 = - \frac{28 r_+^2}{\kappa_5^2} \alpha_3  +  \CO(\alpha_i^2) . \label{l3KP}
\end{align}
In Eqs.~\eqref{etaKP2}-\eqref{l3KP}, $r_+$  is the location of the event horizon in the full black 
brane solution involving all three $\alpha_i$ corrections. The results for $\tp$ and $\kappa$ were previously derived in \cite{Banerjee:2010zd} and are in agreement with our Eqs.~\eqref{etatpKP} and \eqref{kappaKP}. 
The expressions for $\lambda_{1}$, $\lambda_2$ and $\lambda_3$ are new.

Finally, by using the expressions \eqref{etatpKP}, \eqref{l1KP} and \eqref{l2KP}, we confirm that the Haack-Yarom relation among 
the second order coefficients is 
satisfied to linear order in $\alpha_i$ for a (hypothetical) holographic liquid dual to five-dimensional gravity with generic curvature 
squared terms given by the action 
\eqref{R2Th}:
\begin{align}
2 \eta \tp - 4 \lambda_1 - \lambda_2 = \CO(\alpha_i^2 ).
\end{align}

\section{Conclusions}
In this paper, we have made an observation that the universal relation (\ref{universal_identity}) among the second order transport coefficients holds not 
only to leading order in conformal liquids with dual gravity description, as suggested by Haack-Yarom theorem, but remains valid to 
next to leading order in $\CN=4$ SYM and in a (hypothetical) fluid dual to five-dimensional gravity with generic curvature squared terms 
(in particular, Gauss-Bonnet gravity). It is not clear to us whether this result can be generalized to an arbitrary 
higher-derivative correction to Einstein-Hilbert action (to linear order in the corresponding couplings). Such a generalization would follow 
from knowing (from e.g. inequalities obeyed by correlation functions or restrictions imposed by the entropy current) that 
$H(\lambda_{HD})\leq 0$, where $\lambda_{HD}$ is the higher-derivative coupling,  
since then $\lambda_{HD}=0$ would be a maximum of the function $H(\lambda_{HD})$ with $H(0) = 0$ by the Haack-Yarom theorem, and the linear term in the expansion of $H(\lambda_{HD})$ for small 
$\lambda_{HD}$ would necessarily vanish. This is indeed the case for the Gauss-Bonnet gravity, see Eq.~\eqref{SecondOrderUniViolation}. Before looking for a general proof, however, it 
may be useful to check other examples, including other dimensions and charged backgrounds. Weak coupling calculations seem to 
suggest that $H$ is a non-trivial function of the coupling, yet it would be desirable to know this explicitly for a conformal theory 
(e.g. $\CN=4$ SYM) at weak coupling. Also, it may be interesting to generalize the Haack-Yarom theorem to non-conformal holographic liquids\footnote{As shown by Bigazzi and Cotrone \cite{Bigazzi:2010ku}, the relation (\ref{universal_identity})  remains valid in holographic theories where conformality is
broken by a marginally relevant deformation, at leading order in the deformation parameter.}.

The physical significance of the function $H$ is not entirely clear but it might be related to one of the parameters regulating dissipation. The normalized rate of the entropy production in a  conformal fluid near equilibrium is given by \cite{Romatschke:2009kr} 
\begin{align}
\frac{\nabla_a s^a}{s} = \frac{\eta}{2 s T} \sigma_{ab}\sigma^{ab} +\frac{\kappa-2\lambda_1}{4 s T} \sigma_{ab}\sigma^a_c \sigma^{bc} +  \left( \frac{A_1}{2 s} +\frac{\kappa - \eta \tau_\Pi}{2 sT}\right) \sigma_{ab} \left[{}^{\langle} D \sigma^{ab}{}^{\rangle} + \frac{1}{3} \sigma^{ab} \left( \nabla \cdot u\right)\right]\,,
\end{align}
 where the coefficient $A_1$ remains unknown at present. Dissipationless conformal fluids have $\eta =0$, $\kappa = 2 \lambda_1$ and $2\eta \tau_\Pi - 4 \lambda_1 - \lambda_2 =0$ \cite{Bhattacharya:2012zx}, i.e. for such fluids $A_1 = \lambda_2/2T$. Strongly coupled ${\cal N}=4$ SYM is not a dissipationless fluid\footnote{We note that the Gauss-Bonnet fluid dual to (\ref{GBaction}) in the limit $\lgb\rightarrow 1/4$ is not a dissipationless liquid, either: it has vanishing shear viscosity but $\kappa \neq 2 \lambda_1$ and $2\eta \tau_\Pi - 4 \lambda_1 - \lambda_2 \neq 0$ in that limit \cite{Grozdanov:2015asa,Grozdanov:2016fkt}.}, but with $\kappa - 2 \lambda_1 = O(\lambda^{-3/2})$, low viscosity-entropy ratio and $H=0$ it may be not too far from it, comparing especially to the weak coupling limit $\lambda \rightarrow 0$, where 
$\eta / s \sim 1/ \lambda^2 \ln{\lambda^{-1}}$, 
$\lambda_1 \sim T^2/ \lambda^4 \ln^2 \lambda^{-1}$ and $\kappa \sim T^2/ \lambda^2$ \cite{York:2008rr}. This raises the possibility that the near-equilibrium hydrodynamic entropy production is generically suppressed at strong coupling as discussed recently in \cite{Haehl:2014zda}.



\acknowledgments{We would like to thank Sayantani Bhattacharya, 
Guy Moore, Paul Romatschke and Evgeny Shaverin for correspondence, and Alex Buchel, Pavel Kovtun, Rob Myers, Andy O'Bannon, 
Mukund Rangamani, Julian Sonner, Larry Yaffe and Amos Yarom for discussions. This work was 
supported by the European Research Council under the
European Union's Seventh Framework Programme (ERC Grant agreement 307955).}

\bibliography{second_order_transport_relation_2014}

\begin{thebibliography}{61}%
\makeatletter
\providecommand \@ifxundefined [1]{%
 \@ifx{#1\undefined}
}%
\providecommand \@ifnum [1]{%
 \ifnum #1\expandafter \@firstoftwo
 \else \expandafter \@secondoftwo
 \fi
}%
\providecommand \@ifx [1]{%
 \ifx #1\expandafter \@firstoftwo
 \else \expandafter \@secondoftwo
 \fi
}%
\providecommand \natexlab [1]{#1}%
\providecommand \enquote  [1]{``#1''}%
\providecommand \bibnamefont  [1]{#1}%
\providecommand \bibfnamefont [1]{#1}%
\providecommand \citenamefont [1]{#1}%
\providecommand \href@noop [0]{\@secondoftwo}%
\providecommand \href [0]{\begingroup \@sanitize@url \@href}%
\providecommand \@href[1]{\@@startlink{#1}\@@href}%
\providecommand \@@href[1]{\endgroup#1\@@endlink}%
\providecommand \@sanitize@url [0]{\catcode `\\12\catcode `\$12\catcode
  `\&12\catcode `\#12\catcode `\^12\catcode `\_12\catcode `\%12\relax}%
\providecommand \@@startlink[1]{}%
\providecommand \@@endlink[0]{}%
\providecommand \url  [0]{\begingroup\@sanitize@url \@url }%
\providecommand \@url [1]{\endgroup\@href {#1}{\urlprefix }}%
\providecommand \urlprefix  [0]{URL }%
\providecommand \Eprint [0]{\href }%
\providecommand \doibase [0]{http://dx.doi.org/}%
\providecommand \selectlanguage [0]{\@gobble}%
\providecommand \bibinfo  [0]{\@secondoftwo}%
\providecommand \bibfield  [0]{\@secondoftwo}%
\providecommand \translation [1]{[#1]}%
\providecommand \BibitemOpen [0]{}%
\providecommand \bibitemStop [0]{}%
\providecommand \bibitemNoStop [0]{.\EOS\space}%
\providecommand \EOS [0]{\spacefactor3000\relax}%
\providecommand \BibitemShut  [1]{\csname bibitem#1\endcsname}%
\let\auto@bib@innerbib\@empty
\bibitem [{\citenamefont {Forster}(1995)}]{forster}%
  \BibitemOpen
  \bibfield  {author} {\bibinfo {author} {\bibfnamefont {D.}~\bibnamefont
  {Forster}},\ }\href@noop {} {\emph {\bibinfo {title} {{Hydrodynamic
  Fluctuations, Broken Symmetry, and Correlation Functions}}}}\ (\bibinfo
  {publisher} {Westview Press},\ \bibinfo {year} {1995})\BibitemShut {NoStop}%
\bibitem [{\citenamefont {Baier}\ \emph {et~al.}(2008)\citenamefont {Baier},
  \citenamefont {Romatschke}, \citenamefont {Son}, \citenamefont {Starinets},\
  and\ \citenamefont {Stephanov}}]{Baier:2007ix}%
  \BibitemOpen
  \bibfield  {author} {\bibinfo {author} {\bibfnamefont {Rudolf}\ \bibnamefont
  {Baier}}, \bibinfo {author} {\bibfnamefont {Paul}\ \bibnamefont
  {Romatschke}}, \bibinfo {author} {\bibfnamefont {Dam~Thanh}\ \bibnamefont
  {Son}}, \bibinfo {author} {\bibfnamefont {Andrei~O.}\ \bibnamefont
  {Starinets}}, \ and\ \bibinfo {author} {\bibfnamefont {Mikhail~A.}\
  \bibnamefont {Stephanov}},\ }\bibfield  {title} {\enquote {\bibinfo {title}
  {{Relativistic viscous hydrodynamics, conformal invariance, and
  holography}},}\ }\href {\doibase 10.1088/1126-6708/2008/04/100} {\bibfield
  {journal} {\bibinfo  {journal} {JHEP}\ }\textbf {\bibinfo {volume} {0804}},\
  \bibinfo {pages} {100} (\bibinfo {year} {2008})},\ \Eprint
  {http://arxiv.org/abs/0712.2451} {arXiv:0712.2451 [hep-th]} \BibitemShut
  {NoStop}%
\bibitem [{\citenamefont {Son}\ and\ \citenamefont
  {Starinets}(2007)}]{Son:2007vk}%
  \BibitemOpen
  \bibfield  {author} {\bibinfo {author} {\bibfnamefont {Dam~T.}\ \bibnamefont
  {Son}}\ and\ \bibinfo {author} {\bibfnamefont {Andrei~O.}\ \bibnamefont
  {Starinets}},\ }\bibfield  {title} {\enquote {\bibinfo {title} {{Viscosity,
  Black Holes, and Quantum Field Theory}},}\ }\href {\doibase
  10.1146/annurev.nucl.57.090506.123120} {\bibfield  {journal} {\bibinfo
  {journal} {Ann.Rev.Nucl.Part.Sci.}\ }\textbf {\bibinfo {volume} {57}},\
  \bibinfo {pages} {95--118} (\bibinfo {year} {2007})},\ \Eprint
  {http://arxiv.org/abs/0704.0240} {arXiv:0704.0240 [hep-th]} \BibitemShut
  {NoStop}%
\bibitem [{\citenamefont {Schaefer}()}]{Schaefer:2014awa}%
  \BibitemOpen
  \bibfield  {author} {\bibinfo {author} {\bibfnamefont {Thomas}\ \bibnamefont
  {Schaefer}},\ }\bibfield  {title} {\enquote {\bibinfo {title} {{Fluid
  Dynamics and Viscosity in Strongly Correlated Fluids}},}\ }\href@noop {} {\
  }\Eprint {http://arxiv.org/abs/1403.0653} {arXiv:1403.0653 [hep-ph]}
  \BibitemShut {NoStop}%
\bibitem [{\citenamefont {Kovtun}\ and\ \citenamefont
  {Yaffe}(2003)}]{Kovtun:2003vj}%
  \BibitemOpen
  \bibfield  {author} {\bibinfo {author} {\bibfnamefont {Pavel}\ \bibnamefont
  {Kovtun}}\ and\ \bibinfo {author} {\bibfnamefont {Laurence~G.}\ \bibnamefont
  {Yaffe}},\ }\bibfield  {title} {\enquote {\bibinfo {title} {{Hydrodynamic
  fluctuations, long time tails, and supersymmetry}},}\ }\href {\doibase
  10.1103/PhysRevD.68.025007} {\bibfield  {journal} {\bibinfo  {journal}
  {Phys.Rev.}\ }\textbf {\bibinfo {volume} {D68}},\ \bibinfo {pages} {025007}
  (\bibinfo {year} {2003})},\ \Eprint {http://arxiv.org/abs/hep-th/0303010}
  {arXiv:hep-th/0303010 [hep-th]} \BibitemShut {NoStop}%
\bibitem [{\citenamefont {Kovtun}\ \emph {et~al.}(2011)\citenamefont {Kovtun},
  \citenamefont {Moore},\ and\ \citenamefont {Romatschke}}]{Kovtun:2011np}%
  \BibitemOpen
  \bibfield  {author} {\bibinfo {author} {\bibfnamefont {Pavel}\ \bibnamefont
  {Kovtun}}, \bibinfo {author} {\bibfnamefont {Guy~D.}\ \bibnamefont {Moore}},
  \ and\ \bibinfo {author} {\bibfnamefont {Paul}\ \bibnamefont {Romatschke}},\
  }\bibfield  {title} {\enquote {\bibinfo {title} {{The stickiness of sound: An
  absolute lower limit on viscosity and the breakdown of second order
  relativistic hydrodynamics}},}\ }\href {\doibase 10.1103/PhysRevD.84.025006}
  {\bibfield  {journal} {\bibinfo  {journal} {Phys.Rev.}\ }\textbf {\bibinfo
  {volume} {D84}},\ \bibinfo {pages} {025006} (\bibinfo {year} {2011})},\
  \Eprint {http://arxiv.org/abs/1104.1586} {arXiv:1104.1586 [hep-ph]}
  \BibitemShut {NoStop}%
\bibitem [{\citenamefont {Maldacena}(1998)}]{original1}%
  \BibitemOpen
  \bibfield  {author} {\bibinfo {author} {\bibfnamefont {J.~M.}\ \bibnamefont
  {Maldacena}},\ }\bibfield  {title} {\enquote {\bibinfo {title} {{The large N
  limit of superconformal field theories and supergravity}},}\ }\href@noop {}
  {\bibfield  {journal} {\bibinfo  {journal} {Adv.~Theor.~Math.~Phys.}\
  }\textbf {\bibinfo {volume} {2}},\ \bibinfo {pages} {231} (\bibinfo {year}
  {1998})},\ \Eprint {http://arxiv.org/abs/hep-th/9711200}
  {arXiv:hep-th/9711200} \BibitemShut {NoStop}%
\bibitem [{\citenamefont {Gubser}\ \emph
  {et~al.}(1998{\natexlab{a}})\citenamefont {Gubser}, \citenamefont
  {Klebanov},\ and\ \citenamefont {Polyakov}}]{original2}%
  \BibitemOpen
  \bibfield  {author} {\bibinfo {author} {\bibfnamefont {S.~S.}\ \bibnamefont
  {Gubser}}, \bibinfo {author} {\bibfnamefont {I.~R.}\ \bibnamefont
  {Klebanov}}, \ and\ \bibinfo {author} {\bibfnamefont {A.~M.}\ \bibnamefont
  {Polyakov}},\ }\bibfield  {title} {\enquote {\bibinfo {title} {{Gauge theory
  correlators from non-critical string theory}},}\ }\href@noop {} {\bibfield
  {journal} {\bibinfo  {journal} {Phys.~Lett.~B}\ }\textbf {\bibinfo {volume}
  {428}},\ \bibinfo {pages} {105} (\bibinfo {year} {1998}{\natexlab{a}})},\
  \Eprint {http://arxiv.org/abs/hep-th/9802109} {arXiv:hep-th/9802109}
  \BibitemShut {NoStop}%
\bibitem [{\citenamefont {Witten}(1998)}]{original3}%
  \BibitemOpen
  \bibfield  {author} {\bibinfo {author} {\bibfnamefont {E.}~\bibnamefont
  {Witten}},\ }\bibfield  {title} {\enquote {\bibinfo {title} {{Anti de Sitter
  space and holography}},}\ }\href@noop {} {\bibfield  {journal} {\bibinfo
  {journal} {Adv.~Theor.~Math.~Phys.}\ }\textbf {\bibinfo {volume} {2}},\
  \bibinfo {pages} {253} (\bibinfo {year} {1998})},\ \Eprint
  {http://arxiv.org/abs/hep-th/9802150} {arXiv:hep-th/9802150} \BibitemShut
  {NoStop}%
\bibitem [{\citenamefont {Aharony}\ \emph {et~al.}(2000)\citenamefont
  {Aharony}, \citenamefont {Gubser}, \citenamefont {Maldacena}, \citenamefont
  {Ooguri},\ and\ \citenamefont {Oz}}]{original4}%
  \BibitemOpen
  \bibfield  {author} {\bibinfo {author} {\bibfnamefont {O.}~\bibnamefont
  {Aharony}}, \bibinfo {author} {\bibfnamefont {S.~S.}\ \bibnamefont {Gubser}},
  \bibinfo {author} {\bibfnamefont {J.}~\bibnamefont {Maldacena}}, \bibinfo
  {author} {\bibfnamefont {H.}~\bibnamefont {Ooguri}}, \ and\ \bibinfo {author}
  {\bibfnamefont {Y.}~\bibnamefont {Oz}},\ }\bibfield  {title} {\enquote
  {\bibinfo {title} {{Large N field theories, string theory and gravity}},}\
  }\href@noop {} {\bibfield  {journal} {\bibinfo  {journal} {Phys.~Rept.}\
  }\textbf {\bibinfo {volume} {323}},\ \bibinfo {pages} {183} (\bibinfo {year}
  {2000})},\ \Eprint {http://arxiv.org/abs/hep-th/9905111}
  {arXiv:hep-th/9905111} \BibitemShut {NoStop}%
\bibitem [{\citenamefont {Kovtun}\ \emph {et~al.}(2005)\citenamefont {Kovtun},
  \citenamefont {Son},\ and\ \citenamefont {Starinets}}]{Kovtun:2004de}%
  \BibitemOpen
  \bibfield  {author} {\bibinfo {author} {\bibfnamefont {P.}~\bibnamefont
  {Kovtun}}, \bibinfo {author} {\bibfnamefont {Dam~T.}\ \bibnamefont {Son}}, \
  and\ \bibinfo {author} {\bibfnamefont {Andrei~O.}\ \bibnamefont
  {Starinets}},\ }\bibfield  {title} {\enquote {\bibinfo {title} {{Viscosity in
  strongly interacting quantum field theories from black hole physics}},}\
  }\href {\doibase 10.1103/PhysRevLett.94.111601} {\bibfield  {journal}
  {\bibinfo  {journal} {Phys.Rev.Lett.}\ }\textbf {\bibinfo {volume} {94}},\
  \bibinfo {pages} {111601} (\bibinfo {year} {2005})},\ \Eprint
  {http://arxiv.org/abs/hep-th/0405231} {arXiv:hep-th/0405231 [hep-th]}
  \BibitemShut {NoStop}%
\bibitem [{\citenamefont {Buchel}(2005)}]{Buchel:2004qq}%
  \BibitemOpen
  \bibfield  {author} {\bibinfo {author} {\bibfnamefont {Alex}\ \bibnamefont
  {Buchel}},\ }\bibfield  {title} {\enquote {\bibinfo {title} {{On universality
  of stress-energy tensor correlation functions in supergravity}},}\ }\href
  {\doibase 10.1016/j.physletb.2005.01.052} {\bibfield  {journal} {\bibinfo
  {journal} {Phys.Lett.}\ }\textbf {\bibinfo {volume} {B609}},\ \bibinfo
  {pages} {392--401} (\bibinfo {year} {2005})},\ \Eprint
  {http://arxiv.org/abs/hep-th/0408095} {arXiv:hep-th/0408095 [hep-th]}
  \BibitemShut {NoStop}%
\bibitem [{\citenamefont {Kovtun}\ \emph {et~al.}(2003)\citenamefont {Kovtun},
  \citenamefont {Son},\ and\ \citenamefont {Starinets}}]{Kovtun:2003wp}%
  \BibitemOpen
  \bibfield  {author} {\bibinfo {author} {\bibfnamefont {Pavel}\ \bibnamefont
  {Kovtun}}, \bibinfo {author} {\bibfnamefont {Dam~T.}\ \bibnamefont {Son}}, \
  and\ \bibinfo {author} {\bibfnamefont {Andrei~O.}\ \bibnamefont
  {Starinets}},\ }\bibfield  {title} {\enquote {\bibinfo {title} {{Holography
  and hydrodynamics: Diffusion on stretched horizons}},}\ }\href {\doibase
  10.1088/1126-6708/2003/10/064} {\bibfield  {journal} {\bibinfo  {journal}
  {JHEP}\ }\textbf {\bibinfo {volume} {0310}},\ \bibinfo {pages} {064}
  (\bibinfo {year} {2003})},\ \Eprint {http://arxiv.org/abs/hep-th/0309213}
  {arXiv:hep-th/0309213 [hep-th]} \BibitemShut {NoStop}%
\bibitem [{\citenamefont {Buchel}\ and\ \citenamefont
  {Liu}(2004)}]{Buchel:2003tz}%
  \BibitemOpen
  \bibfield  {author} {\bibinfo {author} {\bibfnamefont {Alex}\ \bibnamefont
  {Buchel}}\ and\ \bibinfo {author} {\bibfnamefont {James~T.}\ \bibnamefont
  {Liu}},\ }\bibfield  {title} {\enquote {\bibinfo {title} {{Universality of
  the shear viscosity in supergravity}},}\ }\href {\doibase
  10.1103/PhysRevLett.93.090602} {\bibfield  {journal} {\bibinfo  {journal}
  {Phys.Rev.Lett.}\ }\textbf {\bibinfo {volume} {93}},\ \bibinfo {pages}
  {090602} (\bibinfo {year} {2004})},\ \Eprint
  {http://arxiv.org/abs/hep-th/0311175} {arXiv:hep-th/0311175 [hep-th]}
  \BibitemShut {NoStop}%
\bibitem [{\citenamefont {Starinets}(2009)}]{Starinets:2008fb}%
  \BibitemOpen
  \bibfield  {author} {\bibinfo {author} {\bibfnamefont {Andrei~O.}\
  \bibnamefont {Starinets}},\ }\bibfield  {title} {\enquote {\bibinfo {title}
  {{Quasinormal spectrum and the black hole membrane paradigm}},}\ }\href
  {\doibase 10.1016/j.physletb.2008.11.028} {\bibfield  {journal} {\bibinfo
  {journal} {Phys.Lett.}\ }\textbf {\bibinfo {volume} {B670}},\ \bibinfo
  {pages} {442--445} (\bibinfo {year} {2009})},\ \Eprint
  {http://arxiv.org/abs/0806.3797} {arXiv:0806.3797 [hep-th]} \BibitemShut
  {NoStop}%
\bibitem [{\citenamefont {Cremonini}(2011)}]{Cremonini:2011iq}%
  \BibitemOpen
  \bibfield  {author} {\bibinfo {author} {\bibfnamefont {S.}~\bibnamefont
  {Cremonini}},\ }\bibfield  {title} {\enquote {\bibinfo {title} {{The Shear
  Viscosity to Entropy Ratio: A Status Report}},}\ }\href {\doibase
  10.1142/S0217984911027315} {\bibfield  {journal} {\bibinfo  {journal}
  {Mod.Phys.Lett.}\ }\textbf {\bibinfo {volume} {B25}},\ \bibinfo {pages}
  {1867--1888} (\bibinfo {year} {2011})},\ \Eprint
  {http://arxiv.org/abs/1108.0677} {arXiv:1108.0677 [hep-th]} \BibitemShut
  {NoStop}%
\bibitem [{\citenamefont {Buchel}\ \emph {et~al.}(2005)\citenamefont {Buchel},
  \citenamefont {Liu},\ and\ \citenamefont {Starinets}}]{Buchel:2004di}%
  \BibitemOpen
  \bibfield  {author} {\bibinfo {author} {\bibfnamefont {Alex}\ \bibnamefont
  {Buchel}}, \bibinfo {author} {\bibfnamefont {James~T.}\ \bibnamefont {Liu}},
  \ and\ \bibinfo {author} {\bibfnamefont {Andrei~O.}\ \bibnamefont
  {Starinets}},\ }\bibfield  {title} {\enquote {\bibinfo {title} {{Coupling
  constant dependence of the shear viscosity in N=4 supersymmetric Yang-Mills
  theory}},}\ }\href {\doibase 10.1016/j.nuclphysb.2004.11.055} {\bibfield
  {journal} {\bibinfo  {journal} {Nucl.Phys.}\ }\textbf {\bibinfo {volume}
  {B707}},\ \bibinfo {pages} {56--68} (\bibinfo {year} {2005})},\ \Eprint
  {http://arxiv.org/abs/hep-th/0406264} {arXiv:hep-th/0406264 [hep-th]}
  \BibitemShut {NoStop}%
\bibitem [{\citenamefont {Buchel}(2008{\natexlab{a}})}]{Buchel:2008sh}%
  \BibitemOpen
  \bibfield  {author} {\bibinfo {author} {\bibfnamefont {Alex}\ \bibnamefont
  {Buchel}},\ }\bibfield  {title} {\enquote {\bibinfo {title} {{Resolving
  disagreement for eta/s in a CFT plasma at finite coupling}},}\ }\href
  {\doibase 10.1016/j.nuclphysb.2008.05.024} {\bibfield  {journal} {\bibinfo
  {journal} {Nucl.Phys.}\ }\textbf {\bibinfo {volume} {B803}},\ \bibinfo
  {pages} {166--170} (\bibinfo {year} {2008}{\natexlab{a}})},\ \Eprint
  {http://arxiv.org/abs/0805.2683} {arXiv:0805.2683 [hep-th]} \BibitemShut
  {NoStop}%
\bibitem [{\citenamefont {Kats}\ and\ \citenamefont
  {Petrov}(2009)}]{Kats:2007mq}%
  \BibitemOpen
  \bibfield  {author} {\bibinfo {author} {\bibfnamefont {Yevgeny}\ \bibnamefont
  {Kats}}\ and\ \bibinfo {author} {\bibfnamefont {Pavel}\ \bibnamefont
  {Petrov}},\ }\bibfield  {title} {\enquote {\bibinfo {title} {{Effect of
  curvature squared corrections in AdS on the viscosity of the dual gauge
  theory}},}\ }\href {\doibase 10.1088/1126-6708/2009/01/044} {\bibfield
  {journal} {\bibinfo  {journal} {JHEP}\ }\textbf {\bibinfo {volume} {0901}},\
  \bibinfo {pages} {044} (\bibinfo {year} {2009})},\ \Eprint
  {http://arxiv.org/abs/0712.0743} {arXiv:0712.0743 [hep-th]} \BibitemShut
  {NoStop}%
\bibitem [{\citenamefont {Brigante}\ \emph
  {et~al.}(2008{\natexlab{a}})\citenamefont {Brigante}, \citenamefont {Liu},
  \citenamefont {Myers}, \citenamefont {Shenker},\ and\ \citenamefont
  {Yaida}}]{Brigante:2007nu}%
  \BibitemOpen
  \bibfield  {author} {\bibinfo {author} {\bibfnamefont {Mauro}\ \bibnamefont
  {Brigante}}, \bibinfo {author} {\bibfnamefont {Hong}\ \bibnamefont {Liu}},
  \bibinfo {author} {\bibfnamefont {Robert~C.}\ \bibnamefont {Myers}}, \bibinfo
  {author} {\bibfnamefont {Stephen}\ \bibnamefont {Shenker}}, \ and\ \bibinfo
  {author} {\bibfnamefont {Sho}\ \bibnamefont {Yaida}},\ }\bibfield  {title}
  {\enquote {\bibinfo {title} {{Viscosity Bound Violation in Higher Derivative
  Gravity}},}\ }\href {\doibase 10.1103/PhysRevD.77.126006} {\bibfield
  {journal} {\bibinfo  {journal} {Phys.Rev.}\ }\textbf {\bibinfo {volume}
  {D77}},\ \bibinfo {pages} {126006} (\bibinfo {year} {2008}{\natexlab{a}})},\
  \Eprint {http://arxiv.org/abs/0712.0805} {arXiv:0712.0805 [hep-th]}
  \BibitemShut {NoStop}%
\bibitem [{\citenamefont {Brigante}\ \emph
  {et~al.}(2008{\natexlab{b}})\citenamefont {Brigante}, \citenamefont {Liu},
  \citenamefont {Myers}, \citenamefont {Shenker},\ and\ \citenamefont
  {Yaida}}]{Brigante:2008gz}%
  \BibitemOpen
  \bibfield  {author} {\bibinfo {author} {\bibfnamefont {Mauro}\ \bibnamefont
  {Brigante}}, \bibinfo {author} {\bibfnamefont {Hong}\ \bibnamefont {Liu}},
  \bibinfo {author} {\bibfnamefont {Robert~C.}\ \bibnamefont {Myers}}, \bibinfo
  {author} {\bibfnamefont {Stephen}\ \bibnamefont {Shenker}}, \ and\ \bibinfo
  {author} {\bibfnamefont {Sho}\ \bibnamefont {Yaida}},\ }\bibfield  {title}
  {\enquote {\bibinfo {title} {{The Viscosity Bound and Causality
  Violation}},}\ }\href {\doibase 10.1103/PhysRevLett.100.191601} {\bibfield
  {journal} {\bibinfo  {journal} {Phys.Rev.Lett.}\ }\textbf {\bibinfo {volume}
  {100}},\ \bibinfo {pages} {191601} (\bibinfo {year} {2008}{\natexlab{b}})},\
  \Eprint {http://arxiv.org/abs/0802.3318} {arXiv:0802.3318 [hep-th]}
  \BibitemShut {NoStop}%
\bibitem [{\citenamefont {Buchel}\ and\ \citenamefont
  {Myers}(2009)}]{Buchel:2009tt}%
  \BibitemOpen
  \bibfield  {author} {\bibinfo {author} {\bibfnamefont {Alex}\ \bibnamefont
  {Buchel}}\ and\ \bibinfo {author} {\bibfnamefont {Robert~C.}\ \bibnamefont
  {Myers}},\ }\bibfield  {title} {\enquote {\bibinfo {title} {{Causality of
  Holographic Hydrodynamics}},}\ }\href {\doibase
  10.1088/1126-6708/2009/08/016} {\bibfield  {journal} {\bibinfo  {journal}
  {JHEP}\ }\textbf {\bibinfo {volume} {0908}},\ \bibinfo {pages} {016}
  (\bibinfo {year} {2009})},\ \Eprint {http://arxiv.org/abs/0906.2922}
  {arXiv:0906.2922 [hep-th]} \BibitemShut {NoStop}%
\bibitem [{\citenamefont {Buchel}\ \emph {et~al.}(2010)\citenamefont {Buchel},
  \citenamefont {Escobedo}, \citenamefont {Myers}, \citenamefont {Paulos},
  \citenamefont {Sinha},\ and\ \citenamefont {Smolkin}}]{Buchel:2009sk}%
  \BibitemOpen
  \bibfield  {author} {\bibinfo {author} {\bibfnamefont {Alex}\ \bibnamefont
  {Buchel}}, \bibinfo {author} {\bibfnamefont {Jorge}\ \bibnamefont
  {Escobedo}}, \bibinfo {author} {\bibfnamefont {Robert~C.}\ \bibnamefont
  {Myers}}, \bibinfo {author} {\bibfnamefont {Miguel~F.}\ \bibnamefont
  {Paulos}}, \bibinfo {author} {\bibfnamefont {Aninda}\ \bibnamefont {Sinha}},
  \ and\ \bibinfo {author} {\bibfnamefont {Michael}\ \bibnamefont {Smolkin}},\
  }\bibfield  {title} {\enquote {\bibinfo {title} {{Holographic GB gravity in
  arbitrary dimensions}},}\ }\href {\doibase 10.1007/JHEP03(2010)111}
  {\bibfield  {journal} {\bibinfo  {journal} {JHEP}\ }\textbf {\bibinfo
  {volume} {1003}},\ \bibinfo {pages} {111} (\bibinfo {year} {2010})},\ \Eprint
  {http://arxiv.org/abs/0911.4257} {arXiv:0911.4257 [hep-th]} \BibitemShut
  {NoStop}%
\bibitem [{\citenamefont {Buchel}\ and\ \citenamefont
  {Cremonini}(2010)}]{Buchel:2010wf}%
  \BibitemOpen
  \bibfield  {author} {\bibinfo {author} {\bibfnamefont {Alex}\ \bibnamefont
  {Buchel}}\ and\ \bibinfo {author} {\bibfnamefont {Sera}\ \bibnamefont
  {Cremonini}},\ }\bibfield  {title} {\enquote {\bibinfo {title} {{Viscosity
  Bound and Causality in Superfluid Plasma}},}\ }\href {\doibase
  10.1007/JHEP10(2010)026} {\bibfield  {journal} {\bibinfo  {journal} {JHEP}\
  }\textbf {\bibinfo {volume} {1010}},\ \bibinfo {pages} {026} (\bibinfo {year}
  {2010})},\ \Eprint {http://arxiv.org/abs/1007.2963} {arXiv:1007.2963
  [hep-th]} \BibitemShut {NoStop}%
\bibitem [{\citenamefont {Camanho}\ \emph {et~al.}()\citenamefont {Camanho},
  \citenamefont {Edelstein}, \citenamefont {Maldacena},\ and\ \citenamefont
  {Zhiboedov}}]{Camanho:2014apa}%
  \BibitemOpen
  \bibfield  {author} {\bibinfo {author} {\bibfnamefont {Xian~O.}\ \bibnamefont
  {Camanho}}, \bibinfo {author} {\bibfnamefont {Jose~D.}\ \bibnamefont
  {Edelstein}}, \bibinfo {author} {\bibfnamefont {Juan}\ \bibnamefont
  {Maldacena}}, \ and\ \bibinfo {author} {\bibfnamefont {Alexander}\
  \bibnamefont {Zhiboedov}},\ }\bibfield  {title} {\enquote {\bibinfo {title}
  {{Causality Constraints on Corrections to the Graviton Three-Point
  Coupling}},}\ }\href@noop {} {\ }\Eprint {http://arxiv.org/abs/1407.5597}
  {arXiv:1407.5597 [hep-th]} \BibitemShut {NoStop}%
\bibitem [{\citenamefont {Benincasa}\ \emph {et~al.}(2006)\citenamefont
  {Benincasa}, \citenamefont {Buchel},\ and\ \citenamefont
  {Starinets}}]{Benincasa:2005iv}%
  \BibitemOpen
  \bibfield  {author} {\bibinfo {author} {\bibfnamefont {Paolo}\ \bibnamefont
  {Benincasa}}, \bibinfo {author} {\bibfnamefont {Alex}\ \bibnamefont
  {Buchel}}, \ and\ \bibinfo {author} {\bibfnamefont {Andrei~O.}\ \bibnamefont
  {Starinets}},\ }\bibfield  {title} {\enquote {\bibinfo {title} {{Sound waves
  in strongly coupled non-conformal gauge theory plasma}},}\ }\href {\doibase
  10.1016/j.nuclphysb.2005.11.005} {\bibfield  {journal} {\bibinfo  {journal}
  {Nucl.Phys.}\ }\textbf {\bibinfo {volume} {B733}},\ \bibinfo {pages}
  {160--187} (\bibinfo {year} {2006})},\ \Eprint
  {http://arxiv.org/abs/hep-th/0507026} {arXiv:hep-th/0507026 [hep-th]}
  \BibitemShut {NoStop}%
\bibitem [{\citenamefont {Benincasa}\ and\ \citenamefont
  {Buchel}(2006)}]{Benincasa:2005qc}%
  \BibitemOpen
  \bibfield  {author} {\bibinfo {author} {\bibfnamefont {Paolo}\ \bibnamefont
  {Benincasa}}\ and\ \bibinfo {author} {\bibfnamefont {Alex}\ \bibnamefont
  {Buchel}},\ }\bibfield  {title} {\enquote {\bibinfo {title} {{Transport
  properties of N=4 supersymmetric Yang-Mills theory at finite coupling}},}\
  }\href {\doibase 10.1088/1126-6708/2006/01/103} {\bibfield  {journal}
  {\bibinfo  {journal} {JHEP}\ }\textbf {\bibinfo {volume} {0601}},\ \bibinfo
  {pages} {103} (\bibinfo {year} {2006})},\ \Eprint
  {http://arxiv.org/abs/hep-th/0510041} {arXiv:hep-th/0510041 [hep-th]}
  \BibitemShut {NoStop}%
\bibitem [{\citenamefont {Mas}\ and\ \citenamefont
  {Tarrio}(2007)}]{Mas:2007ng}%
  \BibitemOpen
  \bibfield  {author} {\bibinfo {author} {\bibfnamefont {Javier}\ \bibnamefont
  {Mas}}\ and\ \bibinfo {author} {\bibfnamefont {Javier}\ \bibnamefont
  {Tarrio}},\ }\bibfield  {title} {\enquote {\bibinfo {title} {{Hydrodynamics
  from the Dp-brane}},}\ }\href {\doibase 10.1088/1126-6708/2007/05/036}
  {\bibfield  {journal} {\bibinfo  {journal} {JHEP}\ }\textbf {\bibinfo
  {volume} {0705}},\ \bibinfo {pages} {036} (\bibinfo {year} {2007})},\ \Eprint
  {http://arxiv.org/abs/hep-th/0703093} {arXiv:hep-th/0703093 [HEP-TH]}
  \BibitemShut {NoStop}%
\bibitem [{\citenamefont {Buchel}(2008{\natexlab{b}})}]{Buchel:2007mf}%
  \BibitemOpen
  \bibfield  {author} {\bibinfo {author} {\bibfnamefont {A.}~\bibnamefont
  {Buchel}},\ }\bibfield  {title} {\enquote {\bibinfo {title} {{Bulk viscosity
  of gauge theory plasma at strong coupling}},}\ }\href {\doibase
  10.1016/j.physletb.2008.03.069} {\bibfield  {journal} {\bibinfo  {journal}
  {Phys.Lett.}\ }\textbf {\bibinfo {volume} {B663}},\ \bibinfo {pages}
  {286--289} (\bibinfo {year} {2008}{\natexlab{b}})},\ \Eprint
  {http://arxiv.org/abs/0708.3459} {arXiv:0708.3459 [hep-th]} \BibitemShut
  {NoStop}%
\bibitem [{\citenamefont {Buchel}(2012)}]{Buchel:2011uj}%
  \BibitemOpen
  \bibfield  {author} {\bibinfo {author} {\bibfnamefont {A.}~\bibnamefont
  {Buchel}},\ }\bibfield  {title} {\enquote {\bibinfo {title} {{Violation of
  the holographic bulk viscosity bound}},}\ }\href {\doibase
  10.1103/PhysRevD.85.066004} {\bibfield  {journal} {\bibinfo  {journal}
  {Phys.Rev.}\ }\textbf {\bibinfo {volume} {D85}},\ \bibinfo {pages} {066004}
  (\bibinfo {year} {2012})},\ \Eprint {http://arxiv.org/abs/1110.0063}
  {arXiv:1110.0063 [hep-th]} \BibitemShut {NoStop}%
\bibitem [{\citenamefont {Erdmenger}\ \emph {et~al.}(2009)\citenamefont
  {Erdmenger}, \citenamefont {Haack}, \citenamefont {Kaminski},\ and\
  \citenamefont {Yarom}}]{Erdmenger:2008rm}%
  \BibitemOpen
  \bibfield  {author} {\bibinfo {author} {\bibfnamefont {Johanna}\ \bibnamefont
  {Erdmenger}}, \bibinfo {author} {\bibfnamefont {Michael}\ \bibnamefont
  {Haack}}, \bibinfo {author} {\bibfnamefont {Matthias}\ \bibnamefont
  {Kaminski}}, \ and\ \bibinfo {author} {\bibfnamefont {Amos}\ \bibnamefont
  {Yarom}},\ }\bibfield  {title} {\enquote {\bibinfo {title} {{Fluid dynamics
  of R-charged black holes}},}\ }\href {\doibase 10.1088/1126-6708/2009/01/055}
  {\bibfield  {journal} {\bibinfo  {journal} {JHEP}\ }\textbf {\bibinfo
  {volume} {0901}},\ \bibinfo {pages} {055} (\bibinfo {year} {2009})},\ \Eprint
  {http://arxiv.org/abs/0809.2488} {arXiv:0809.2488 [hep-th]} \BibitemShut
  {NoStop}%
\bibitem [{\citenamefont {Haack}\ and\ \citenamefont
  {Yarom}(2009)}]{Haack:2008xx}%
  \BibitemOpen
  \bibfield  {author} {\bibinfo {author} {\bibfnamefont {Michael}\ \bibnamefont
  {Haack}}\ and\ \bibinfo {author} {\bibfnamefont {Amos}\ \bibnamefont
  {Yarom}},\ }\bibfield  {title} {\enquote {\bibinfo {title} {{Universality of
  second order transport coefficients from the gauge-string duality}},}\ }\href
  {\doibase 10.1016/j.nuclphysb.2008.12.028} {\bibfield  {journal} {\bibinfo
  {journal} {Nucl.Phys.}\ }\textbf {\bibinfo {volume} {B813}},\ \bibinfo
  {pages} {140--155} (\bibinfo {year} {2009})},\ \Eprint
  {http://arxiv.org/abs/0811.1794} {arXiv:0811.1794 [hep-th]} \BibitemShut
  {NoStop}%
\bibitem [{\citenamefont {York}\ and\ \citenamefont
  {Moore}(2009)}]{York:2008rr}%
  \BibitemOpen
  \bibfield  {author} {\bibinfo {author} {\bibfnamefont {Mark~Abraao}\
  \bibnamefont {York}}\ and\ \bibinfo {author} {\bibfnamefont {Guy~D.}\
  \bibnamefont {Moore}},\ }\bibfield  {title} {\enquote {\bibinfo {title}
  {{Second order hydrodynamic coefficients from kinetic theory}},}\ }\href
  {\doibase 10.1103/PhysRevD.79.054011} {\bibfield  {journal} {\bibinfo
  {journal} {Phys.Rev.}\ }\textbf {\bibinfo {volume} {D79}},\ \bibinfo {pages}
  {054011} (\bibinfo {year} {2009})},\ \Eprint {http://arxiv.org/abs/0811.0729}
  {arXiv:0811.0729 [hep-ph]} \BibitemShut {NoStop}%
\bibitem [{\citenamefont {Betz}\ \emph {et~al.}(2009)\citenamefont {Betz},
  \citenamefont {Henkel},\ and\ \citenamefont {Rischke}}]{Betz:2008me}%
  \BibitemOpen
  \bibfield  {author} {\bibinfo {author} {\bibfnamefont {B.}~\bibnamefont
  {Betz}}, \bibinfo {author} {\bibfnamefont {D.}~\bibnamefont {Henkel}}, \ and\
  \bibinfo {author} {\bibfnamefont {D.H.}\ \bibnamefont {Rischke}},\ }\bibfield
   {title} {\enquote {\bibinfo {title} {{From kinetic theory to dissipative
  fluid dynamics}},}\ }\href {\doibase 10.1016/j.ppnp.2008.12.018} {\bibfield
  {journal} {\bibinfo  {journal} {Prog.Part.Nucl.Phys.}\ }\textbf {\bibinfo
  {volume} {62}},\ \bibinfo {pages} {556--561} (\bibinfo {year} {2009})},\
  \Eprint {http://arxiv.org/abs/0812.1440} {arXiv:0812.1440 [nucl-th]}
  \BibitemShut {NoStop}%
\bibitem [{\citenamefont {Shaverin}\ and\ \citenamefont
  {Yarom}(2013)}]{Shaverin:2012kv}%
  \BibitemOpen
  \bibfield  {author} {\bibinfo {author} {\bibfnamefont {Evgeny}\ \bibnamefont
  {Shaverin}}\ and\ \bibinfo {author} {\bibfnamefont {Amos}\ \bibnamefont
  {Yarom}},\ }\bibfield  {title} {\enquote {\bibinfo {title} {{Universality of
  second order transport in Gauss-Bonnet gravity}},}\ }\href {\doibase
  10.1007/JHEP04(2013)013} {\bibfield  {journal} {\bibinfo  {journal} {JHEP}\
  }\textbf {\bibinfo {volume} {1304}},\ \bibinfo {pages} {013} (\bibinfo {year}
  {2013})},\ \Eprint {http://arxiv.org/abs/1211.1979} {arXiv:1211.1979
  [hep-th]} \BibitemShut {NoStop}%
\bibitem [{\citenamefont {Grozdanov}\ and\ \citenamefont
  {Starinets}(2015)}]{Grozdanov:2015asa}%
  \BibitemOpen
  \bibfield  {author} {\bibinfo {author} {\bibfnamefont {S.}~\bibnamefont
  {Grozdanov}}\ and\ \bibinfo {author} {\bibfnamefont {A.O.}\ \bibnamefont
  {Starinets}},\ }\bibfield  {title} {\enquote {\bibinfo {title}
  {{Zero-viscosity limit in a holographic Gauss-Bonnet liquid}},}\ }\href
  {\doibase 10.1007/s11232-015-0245-7} {\bibfield  {journal} {\bibinfo
  {journal} {Theor.Math.Phys.}\ }\textbf {\bibinfo {volume} {182}},\ \bibinfo
  {pages} {61--73} (\bibinfo {year} {2015})}\BibitemShut {NoStop}%
\bibitem [{\citenamefont {Grozdanov}\ and\ \citenamefont
  {Starinets}(2016)}]{Grozdanov:2016fkt}%
  \BibitemOpen
  \bibfield  {author} {\bibinfo {author} {\bibfnamefont {Saso}\ \bibnamefont
  {Grozdanov}}\ and\ \bibinfo {author} {\bibfnamefont {Andrei~O.}\ \bibnamefont
  {Starinets}},\ }\bibfield  {title} {\enquote {\bibinfo {title} {{Second-order
  transport, quasinormal modes and zero-viscosity limit in the Gauss-Bonnet
  holographic fluid}},}\ }\href@noop {} {\  (\bibinfo {year} {2016})},\ \Eprint
  {http://arxiv.org/abs/1611.07053} {arXiv:1611.07053 [hep-th]} \BibitemShut
  {NoStop}%
\bibitem [{\citenamefont {Shaverin}(2013)}]{Shaverin-thesis}%
  \BibitemOpen
  \bibfield  {author} {\bibinfo {author} {\bibfnamefont {Evgeny}\ \bibnamefont
  {Shaverin}},\ }\bibfield  {title} {\enquote {\bibinfo {title} {Second order
  hydrodynamics via ads/cft},}\ }\href@noop {} {\bibfield  {journal} {\bibinfo
  {journal} {Master of Science Thesis, Technion - Israel Institute of
  Technology, Haifa}\ } (\bibinfo {year} {2013})}\BibitemShut {NoStop}%
\bibitem [{\citenamefont {Shaverin}(2015)}]{Shaverin:2015vda}%
  \BibitemOpen
  \bibfield  {author} {\bibinfo {author} {\bibfnamefont {Evgeny}\ \bibnamefont
  {Shaverin}},\ }\bibfield  {title} {\enquote {\bibinfo {title} {{A breakdown
  of a universal hydrodynamic relation in Gauss-Bonnet gravity}},}\ }\href@noop
  {} {\  (\bibinfo {year} {2015})},\ \Eprint {http://arxiv.org/abs/1509.05418}
  {arXiv:1509.05418 [hep-th]} \BibitemShut {NoStop}%
\bibitem [{\citenamefont {Policastro}\ \emph {et~al.}(2001)\citenamefont
  {Policastro}, \citenamefont {Son},\ and\ \citenamefont
  {Starinets}}]{Policastro:2001yc}%
  \BibitemOpen
  \bibfield  {author} {\bibinfo {author} {\bibfnamefont {G.}~\bibnamefont
  {Policastro}}, \bibinfo {author} {\bibfnamefont {Dam~T.}\ \bibnamefont
  {Son}}, \ and\ \bibinfo {author} {\bibfnamefont {Andrei~O.}\ \bibnamefont
  {Starinets}},\ }\bibfield  {title} {\enquote {\bibinfo {title} {{The Shear
  viscosity of strongly coupled N=4 supersymmetric Yang-Mills plasma}},}\
  }\href {\doibase 10.1103/PhysRevLett.87.081601} {\bibfield  {journal}
  {\bibinfo  {journal} {Phys.Rev.Lett.}\ }\textbf {\bibinfo {volume} {87}},\
  \bibinfo {pages} {081601} (\bibinfo {year} {2001})},\ \Eprint
  {http://arxiv.org/abs/hep-th/0104066} {arXiv:hep-th/0104066 [hep-th]}
  \BibitemShut {NoStop}%
\bibitem [{\citenamefont {Bhattacharyya}\ \emph {et~al.}(2008)\citenamefont
  {Bhattacharyya}, \citenamefont {Hubeny}, \citenamefont {Minwalla},\ and\
  \citenamefont {Rangamani}}]{Bhattacharyya:2008jc}%
  \BibitemOpen
  \bibfield  {author} {\bibinfo {author} {\bibfnamefont {Sayantani}\
  \bibnamefont {Bhattacharyya}}, \bibinfo {author} {\bibfnamefont {Veronika~E}\
  \bibnamefont {Hubeny}}, \bibinfo {author} {\bibfnamefont {Shiraz}\
  \bibnamefont {Minwalla}}, \ and\ \bibinfo {author} {\bibfnamefont {Mukund}\
  \bibnamefont {Rangamani}},\ }\bibfield  {title} {\enquote {\bibinfo {title}
  {{Nonlinear Fluid Dynamics from Gravity}},}\ }\href {\doibase
  10.1088/1126-6708/2008/02/045} {\bibfield  {journal} {\bibinfo  {journal}
  {JHEP}\ }\textbf {\bibinfo {volume} {0802}},\ \bibinfo {pages} {045}
  (\bibinfo {year} {2008})},\ \Eprint {http://arxiv.org/abs/0712.2456}
  {arXiv:0712.2456 [hep-th]} \BibitemShut {NoStop}%
\bibitem [{\citenamefont {Buchel}(2008{\natexlab{c}})}]{Buchel:2008ac}%
  \BibitemOpen
  \bibfield  {author} {\bibinfo {author} {\bibfnamefont {Alex}\ \bibnamefont
  {Buchel}},\ }\bibfield  {title} {\enquote {\bibinfo {title} {{Shear viscosity
  of boost invariant plasma at finite coupling}},}\ }\href {\doibase
  10.1016/j.nuclphysb.2008.03.009} {\bibfield  {journal} {\bibinfo  {journal}
  {Nucl.Phys.}\ }\textbf {\bibinfo {volume} {B802}},\ \bibinfo {pages}
  {281--306} (\bibinfo {year} {2008}{\natexlab{c}})},\ \Eprint
  {http://arxiv.org/abs/0801.4421} {arXiv:0801.4421 [hep-th]} \BibitemShut
  {NoStop}%
\bibitem [{\citenamefont {Buchel}\ and\ \citenamefont
  {Paulos}(2008)}]{Buchel:2008bz}%
  \BibitemOpen
  \bibfield  {author} {\bibinfo {author} {\bibfnamefont {Alex}\ \bibnamefont
  {Buchel}}\ and\ \bibinfo {author} {\bibfnamefont {Miguel}\ \bibnamefont
  {Paulos}},\ }\bibfield  {title} {\enquote {\bibinfo {title} {{Relaxation time
  of a CFT plasma at finite coupling}},}\ }\href {\doibase
  10.1016/j.nuclphysb.2008.07.002} {\bibfield  {journal} {\bibinfo  {journal}
  {Nucl.Phys.}\ }\textbf {\bibinfo {volume} {B805}},\ \bibinfo {pages} {59--71}
  (\bibinfo {year} {2008})},\ \Eprint {http://arxiv.org/abs/0806.0788}
  {arXiv:0806.0788 [hep-th]} \BibitemShut {NoStop}%
\bibitem [{\citenamefont {Buchel}\ and\ \citenamefont
  {Paulos}(2009)}]{Buchel:2008kd}%
  \BibitemOpen
  \bibfield  {author} {\bibinfo {author} {\bibfnamefont {Alex}\ \bibnamefont
  {Buchel}}\ and\ \bibinfo {author} {\bibfnamefont {Miguel}\ \bibnamefont
  {Paulos}},\ }\bibfield  {title} {\enquote {\bibinfo {title} {{Second order
  hydrodynamics of a CFT plasma from boost invariant expansion}},}\ }\href
  {\doibase 10.1016/j.nuclphysb.2008.10.012} {\bibfield  {journal} {\bibinfo
  {journal} {Nucl.Phys.}\ }\textbf {\bibinfo {volume} {B810}},\ \bibinfo
  {pages} {40--65} (\bibinfo {year} {2009})},\ \Eprint
  {http://arxiv.org/abs/0808.1601} {arXiv:0808.1601 [hep-th]} \BibitemShut
  {NoStop}%
\bibitem [{\citenamefont {Saremi}\ and\ \citenamefont
  {Sohrabi}(2011)}]{Saremi:2011nh}%
  \BibitemOpen
  \bibfield  {author} {\bibinfo {author} {\bibfnamefont {Omid}\ \bibnamefont
  {Saremi}}\ and\ \bibinfo {author} {\bibfnamefont {Kiyoumars~A.}\ \bibnamefont
  {Sohrabi}},\ }\bibfield  {title} {\enquote {\bibinfo {title} {{Causal
  three-point functions and nonlinear second-order hydrodynamic coefficients in
  AdS/CFT}},}\ }\href {\doibase 10.1007/JHEP11(2011)147} {\bibfield  {journal}
  {\bibinfo  {journal} {JHEP}\ }\textbf {\bibinfo {volume} {1111}},\ \bibinfo
  {pages} {147} (\bibinfo {year} {2011})},\ \Eprint
  {http://arxiv.org/abs/1105.4870} {arXiv:1105.4870 [hep-th]} \BibitemShut
  {NoStop}%
\bibitem [{\citenamefont {Huot}\ \emph {et~al.}(2007)\citenamefont {Huot},
  \citenamefont {Jeon},\ and\ \citenamefont {Moore}}]{Huot:2006ys}%
  \BibitemOpen
  \bibfield  {author} {\bibinfo {author} {\bibfnamefont {Simon~C.}\
  \bibnamefont {Huot}}, \bibinfo {author} {\bibfnamefont {Sangyong}\
  \bibnamefont {Jeon}}, \ and\ \bibinfo {author} {\bibfnamefont {Guy~D.}\
  \bibnamefont {Moore}},\ }\bibfield  {title} {\enquote {\bibinfo {title}
  {{Shear viscosity in weakly coupled N = 4 super Yang-Mills theory compared to
  QCD}},}\ }\href {\doibase 10.1103/PhysRevLett.98.172303} {\bibfield
  {journal} {\bibinfo  {journal} {Phys.Rev.Lett.}\ }\textbf {\bibinfo {volume}
  {98}},\ \bibinfo {pages} {172303} (\bibinfo {year} {2007})},\ \Eprint
  {http://arxiv.org/abs/hep-ph/0608062} {arXiv:hep-ph/0608062 [hep-ph]}
  \BibitemShut {NoStop}%
\bibitem [{\citenamefont {Buchel}\ \emph {et~al.}(2008)\citenamefont {Buchel},
  \citenamefont {Myers}, \citenamefont {Paulos},\ and\ \citenamefont
  {Sinha}}]{Buchel:2008ae}%
  \BibitemOpen
  \bibfield  {author} {\bibinfo {author} {\bibfnamefont {Alex}\ \bibnamefont
  {Buchel}}, \bibinfo {author} {\bibfnamefont {Robert~C.}\ \bibnamefont
  {Myers}}, \bibinfo {author} {\bibfnamefont {Miguel~F.}\ \bibnamefont
  {Paulos}}, \ and\ \bibinfo {author} {\bibfnamefont {Aninda}\ \bibnamefont
  {Sinha}},\ }\bibfield  {title} {\enquote {\bibinfo {title} {{Universal
  holographic hydrodynamics at finite coupling}},}\ }\href {\doibase
  10.1016/j.physletb.2008.10.003} {\bibfield  {journal} {\bibinfo  {journal}
  {Phys.Lett.}\ }\textbf {\bibinfo {volume} {B669}},\ \bibinfo {pages}
  {364--370} (\bibinfo {year} {2008})},\ \Eprint
  {http://arxiv.org/abs/0808.1837} {arXiv:0808.1837 [hep-th]} \BibitemShut
  {NoStop}%
\bibitem [{\citenamefont {Gubser}\ \emph
  {et~al.}(1998{\natexlab{b}})\citenamefont {Gubser}, \citenamefont
  {Klebanov},\ and\ \citenamefont {Tseytlin}}]{Gubser:1998nz}%
  \BibitemOpen
  \bibfield  {author} {\bibinfo {author} {\bibfnamefont {Steven~S.}\
  \bibnamefont {Gubser}}, \bibinfo {author} {\bibfnamefont {Igor~R.}\
  \bibnamefont {Klebanov}}, \ and\ \bibinfo {author} {\bibfnamefont
  {Arkady~A.}\ \bibnamefont {Tseytlin}},\ }\bibfield  {title} {\enquote
  {\bibinfo {title} {{Coupling constant dependence in the thermodynamics of N=4
  supersymmetric Yang-Mills theory}},}\ }\href {\doibase
  10.1016/S0550-3213(98)00514-8} {\bibfield  {journal} {\bibinfo  {journal}
  {Nucl.Phys.}\ }\textbf {\bibinfo {volume} {B534}},\ \bibinfo {pages}
  {202--222} (\bibinfo {year} {1998}{\natexlab{b}})},\ \Eprint
  {http://arxiv.org/abs/hep-th/9805156} {arXiv:hep-th/9805156 [hep-th]}
  \BibitemShut {NoStop}%
\bibitem [{\citenamefont {Barnes}\ \emph
  {et~al.}(2010{\natexlab{a}})\citenamefont {Barnes}, \citenamefont {Vaman},
  \citenamefont {Wu},\ and\ \citenamefont {Arnold}}]{Barnes:2010jp}%
  \BibitemOpen
  \bibfield  {author} {\bibinfo {author} {\bibfnamefont {Edwin}\ \bibnamefont
  {Barnes}}, \bibinfo {author} {\bibfnamefont {Diana}\ \bibnamefont {Vaman}},
  \bibinfo {author} {\bibfnamefont {Chaolun}\ \bibnamefont {Wu}}, \ and\
  \bibinfo {author} {\bibfnamefont {Peter}\ \bibnamefont {Arnold}},\ }\bibfield
   {title} {\enquote {\bibinfo {title} {{Real-time finite-temperature
  correlators from AdS/CFT}},}\ }\href {\doibase 10.1103/PhysRevD.82.025019}
  {\bibfield  {journal} {\bibinfo  {journal} {Phys.Rev.}\ }\textbf {\bibinfo
  {volume} {D82}},\ \bibinfo {pages} {025019} (\bibinfo {year}
  {2010}{\natexlab{a}})},\ \Eprint {http://arxiv.org/abs/1004.1179}
  {arXiv:1004.1179 [hep-th]} \BibitemShut {NoStop}%
\bibitem [{\citenamefont {Barnes}\ \emph
  {et~al.}(2010{\natexlab{b}})\citenamefont {Barnes}, \citenamefont {Vaman},\
  and\ \citenamefont {Wu}}]{Barnes:2010ev}%
  \BibitemOpen
  \bibfield  {author} {\bibinfo {author} {\bibfnamefont {Edwin}\ \bibnamefont
  {Barnes}}, \bibinfo {author} {\bibfnamefont {Diana}\ \bibnamefont {Vaman}}, \
  and\ \bibinfo {author} {\bibfnamefont {Chaolun}\ \bibnamefont {Wu}},\
  }\bibfield  {title} {\enquote {\bibinfo {title} {{Holographic real-time
  non-relativistic correlators at zero and finite temperature}},}\ }\href
  {\doibase 10.1103/PhysRevD.82.125042} {\bibfield  {journal} {\bibinfo
  {journal} {Phys. Rev.}\ }\textbf {\bibinfo {volume} {D82}},\ \bibinfo {pages}
  {125042} (\bibinfo {year} {2010}{\natexlab{b}})},\ \Eprint
  {http://arxiv.org/abs/1007.1644} {arXiv:1007.1644 [hep-th]} \BibitemShut
  {NoStop}%
\bibitem [{\citenamefont {Arnold}\ and\ \citenamefont
  {Vaman}(2011)}]{Arnold:2011hp}%
  \BibitemOpen
  \bibfield  {author} {\bibinfo {author} {\bibfnamefont {Peter}\ \bibnamefont
  {Arnold}}\ and\ \bibinfo {author} {\bibfnamefont {Diana}\ \bibnamefont
  {Vaman}},\ }\bibfield  {title} {\enquote {\bibinfo {title} {{4-point
  correlators in finite-temperature AdS/CFT: Jet quenching correlations}},}\
  }\href {\doibase 10.1007/JHEP11(2011)033} {\bibfield  {journal} {\bibinfo
  {journal} {JHEP}\ }\textbf {\bibinfo {volume} {1111}},\ \bibinfo {pages}
  {033} (\bibinfo {year} {2011})},\ \Eprint {http://arxiv.org/abs/1109.0040}
  {arXiv:1109.0040 [hep-th]} \BibitemShut {NoStop}%
\bibitem [{\citenamefont {Moore}\ and\ \citenamefont
  {Sohrabi}(2011)}]{Moore:2010bu}%
  \BibitemOpen
  \bibfield  {author} {\bibinfo {author} {\bibfnamefont {Guy~D.}\ \bibnamefont
  {Moore}}\ and\ \bibinfo {author} {\bibfnamefont {Kiyoumars~A.}\ \bibnamefont
  {Sohrabi}},\ }\bibfield  {title} {\enquote {\bibinfo {title} {{Kubo Formulae
  for Second-Order Hydrodynamic Coefficients}},}\ }\href {\doibase
  10.1103/PhysRevLett.106.122302} {\bibfield  {journal} {\bibinfo  {journal}
  {Phys.Rev.Lett.}\ }\textbf {\bibinfo {volume} {106}},\ \bibinfo {pages}
  {122302} (\bibinfo {year} {2011})},\ \Eprint {http://arxiv.org/abs/1007.5333}
  {arXiv:1007.5333 [hep-ph]} \BibitemShut {NoStop}%
\bibitem [{\citenamefont {Arnold}\ \emph {et~al.}(2011)\citenamefont {Arnold},
  \citenamefont {Vaman}, \citenamefont {Wu},\ and\ \citenamefont
  {Xiao}}]{Arnold:2011ja}%
  \BibitemOpen
  \bibfield  {author} {\bibinfo {author} {\bibfnamefont {Peter}\ \bibnamefont
  {Arnold}}, \bibinfo {author} {\bibfnamefont {Diana}\ \bibnamefont {Vaman}},
  \bibinfo {author} {\bibfnamefont {Chaolun}\ \bibnamefont {Wu}}, \ and\
  \bibinfo {author} {\bibfnamefont {Wei}\ \bibnamefont {Xiao}},\ }\bibfield
  {title} {\enquote {\bibinfo {title} {{Second order hydrodynamic coefficients
  from 3-point stress tensor correlators via AdS/CFT}},}\ }\href {\doibase
  10.1007/JHEP10(2011)033} {\bibfield  {journal} {\bibinfo  {journal} {JHEP}\
  }\textbf {\bibinfo {volume} {1110}},\ \bibinfo {pages} {033} (\bibinfo {year}
  {2011})},\ \Eprint {http://arxiv.org/abs/1105.4645} {arXiv:1105.4645
  [hep-th]} \BibitemShut {NoStop}%
\bibitem [{\citenamefont {Schwinger}(1961)}]{Schwinger:1960qe}%
  \BibitemOpen
  \bibfield  {author} {\bibinfo {author} {\bibfnamefont {Julian~S.}\
  \bibnamefont {Schwinger}},\ }\bibfield  {title} {\enquote {\bibinfo {title}
  {{Brownian motion of a quantum oscillator}},}\ }\href {\doibase
  10.1063/1.1703727} {\bibfield  {journal} {\bibinfo  {journal} {J.Math.Phys.}\
  }\textbf {\bibinfo {volume} {2}},\ \bibinfo {pages} {407--432} (\bibinfo
  {year} {1961})}\BibitemShut {NoStop}%
\bibitem [{\citenamefont {Keldysh}(1964)}]{Keldysh:1964ud}%
  \BibitemOpen
  \bibfield  {author} {\bibinfo {author} {\bibfnamefont {Leonid~V.}\
  \bibnamefont {Keldysh}},\ }\bibfield  {title} {\enquote {\bibinfo {title}
  {{Diagram technique for nonequilibrium processes}},}\ }\href@noop {}
  {\bibfield  {journal} {\bibinfo  {journal} {Zh.Eksp.Teor.Fiz.}\ }\textbf
  {\bibinfo {volume} {47}},\ \bibinfo {pages} {1515--1527} (\bibinfo {year}
  {1964})}\BibitemShut {NoStop}%
\bibitem [{\citenamefont {Wang}\ and\ \citenamefont
  {Heinz}(2002)}]{Wang:1998wg}%
  \BibitemOpen
  \bibfield  {author} {\bibinfo {author} {\bibfnamefont {Enke}\ \bibnamefont
  {Wang}}\ and\ \bibinfo {author} {\bibfnamefont {Ulrich~W.}\ \bibnamefont
  {Heinz}},\ }\bibfield  {title} {\enquote {\bibinfo {title} {{A Generalized
  fluctuation dissipation theorem for nonlinear response functions}},}\ }\href
  {\doibase 10.1103/PhysRevD.66.025008} {\bibfield  {journal} {\bibinfo
  {journal} {Phys.Rev.}\ }\textbf {\bibinfo {volume} {D66}},\ \bibinfo {pages}
  {025008} (\bibinfo {year} {2002})},\ \Eprint
  {http://arxiv.org/abs/hep-th/9809016} {arXiv:hep-th/9809016 [hep-th]}
  \BibitemShut {NoStop}%
\bibitem [{\citenamefont {Banerjee}\ and\ \citenamefont
  {Dutta}(2010)}]{Banerjee:2010zd}%
  \BibitemOpen
  \bibfield  {author} {\bibinfo {author} {\bibfnamefont {Nabamita}\
  \bibnamefont {Banerjee}}\ and\ \bibinfo {author} {\bibfnamefont {Suvankar}\
  \bibnamefont {Dutta}},\ }\bibfield  {title} {\enquote {\bibinfo {title}
  {{Nonlinear Hydrodynamics from Flow of Retarded Green's Function}},}\ }\href
  {\doibase 10.1007/JHEP08(2010)041} {\bibfield  {journal} {\bibinfo  {journal}
  {JHEP}\ }\textbf {\bibinfo {volume} {1008}},\ \bibinfo {pages} {041}
  (\bibinfo {year} {2010})},\ \Eprint {http://arxiv.org/abs/1005.2367}
  {arXiv:1005.2367 [hep-th]} \BibitemShut {NoStop}%
\bibitem [{\citenamefont {Bigazzi}\ and\ \citenamefont
  {Cotrone}(2010)}]{Bigazzi:2010ku}%
  \BibitemOpen
  \bibfield  {author} {\bibinfo {author} {\bibfnamefont {Francesco}\
  \bibnamefont {Bigazzi}}\ and\ \bibinfo {author} {\bibfnamefont {Aldo~L.}\
  \bibnamefont {Cotrone}},\ }\bibfield  {title} {\enquote {\bibinfo {title}
  {{An elementary stringy estimate of transport coefficients of large
  temperature QCD}},}\ }\href {\doibase 10.1007/JHEP08(2010)128} {\bibfield
  {journal} {\bibinfo  {journal} {JHEP}\ }\textbf {\bibinfo {volume} {1008}},\
  \bibinfo {pages} {128} (\bibinfo {year} {2010})},\ \Eprint
  {http://arxiv.org/abs/1006.4634} {arXiv:1006.4634 [hep-ph]} \BibitemShut
  {NoStop}%
\bibitem [{\citenamefont {Romatschke}(2010)}]{Romatschke:2009kr}%
  \BibitemOpen
  \bibfield  {author} {\bibinfo {author} {\bibfnamefont {Paul}\ \bibnamefont
  {Romatschke}},\ }\bibfield  {title} {\enquote {\bibinfo {title}
  {{Relativistic Viscous Fluid Dynamics and Non-Equilibrium Entropy}},}\ }\href
  {\doibase 10.1088/0264-9381/27/2/025006} {\bibfield  {journal} {\bibinfo
  {journal} {Class.Quant.Grav.}\ }\textbf {\bibinfo {volume} {27}},\ \bibinfo
  {pages} {025006} (\bibinfo {year} {2010})},\ \Eprint
  {http://arxiv.org/abs/0906.4787} {arXiv:0906.4787 [hep-th]} \BibitemShut
  {NoStop}%
\bibitem [{\citenamefont {Bhattacharya}\ \emph {et~al.}(2013)\citenamefont
  {Bhattacharya}, \citenamefont {Bhattacharyya},\ and\ \citenamefont
  {Rangamani}}]{Bhattacharya:2012zx}%
  \BibitemOpen
  \bibfield  {author} {\bibinfo {author} {\bibfnamefont {Jyotirmoy}\
  \bibnamefont {Bhattacharya}}, \bibinfo {author} {\bibfnamefont {Sayantani}\
  \bibnamefont {Bhattacharyya}}, \ and\ \bibinfo {author} {\bibfnamefont
  {Mukund}\ \bibnamefont {Rangamani}},\ }\bibfield  {title} {\enquote {\bibinfo
  {title} {{Non-dissipative hydrodynamics: Effective actions versus entropy
  current}},}\ }\href {\doibase 10.1007/JHEP02(2013)153} {\bibfield  {journal}
  {\bibinfo  {journal} {JHEP}\ }\textbf {\bibinfo {volume} {1302}},\ \bibinfo
  {pages} {153} (\bibinfo {year} {2013})},\ \Eprint
  {http://arxiv.org/abs/1211.1020} {arXiv:1211.1020 [hep-th]} \BibitemShut
  {NoStop}%
\bibitem [{\citenamefont {Haehl}\ \emph {et~al.}()\citenamefont {Haehl},
  \citenamefont {Loganayagam},\ and\ \citenamefont
  {Rangamani}}]{Haehl:2014zda}%
  \BibitemOpen
  \bibfield  {author} {\bibinfo {author} {\bibfnamefont {Felix~M.}\
  \bibnamefont {Haehl}}, \bibinfo {author} {\bibfnamefont {R.}~\bibnamefont
  {Loganayagam}}, \ and\ \bibinfo {author} {\bibfnamefont {Mukund}\
  \bibnamefont {Rangamani}},\ }\bibfield  {title} {\enquote {\bibinfo {title}
  {{The eightfold way to dissipation}},}\ }\href@noop {} {\ }\Eprint
  {http://arxiv.org/abs/1412.1090} {arXiv:1412.1090 [hep-th]} \BibitemShut
  {NoStop}%
\end{thebibliography}%

\end{document}